\documentstyle[11pt,fleqn,epsf]{article}

\renewcommand{\baselinestretch}{1.2}

\def\fnote#1#2{\begingroup\def\thefootnote{#1}\footnote{#2}\endgroup}
\makeatletter
\def\section{\@startsection {section}{1}{\z@}{3.5ex plus 1ex minus
    .2ex}{2.3ex plus .2ex}{\sc }}
\def\subsection{\@startsection{subsection}{2}{\z@}{3.25ex plus 1ex minus
   .2ex}{1.5ex plus .2ex}{\small \sc }}
\def\appendix{\par\clearpage
  \setcounter{section}{0}
  \setcounter{subsection}{0}
  \@addtoreset{equation}{section}
  \def\@sectname{Appendix~}
  \def\theequation{\thesection.\arabic{equation}}
  \def\thesection{\Alph{section}}}
\makeatother
\catcode`\@=11
\def\prm{\fam \z@}   % to make \rm work in math
\catcode`\@=12
\makeatletter \@addtoreset{equation}{section} \makeatother
\renewcommand{\theequation}{\thesection.\arabic{equation}}
\def\ap#1#2#3{     {\it Ann. Phys. (NY) }{\bf #1} (19#2) #3}

\def\npb#1#2#3{    {\it Nucl. Phys. }{\bf B #1} (19#2) #3}
\def\plb#1#2#3{    {\it Phys. Lett. }{\bf B #1} (19#2) #3}
\def\prd#1#2#3{    {\it Phys. Rev. }{\bf D #1} (19#2) #3}
\def\prep#1#2#3{   {\it Phys. Rep. }{\bf #1} (19#2) #3}
\def\prl#1#2#3{    {\it Phys. Rev. Lett. }{\bf #1} (19#2) #3}

\def\zpc#1#2#3{    {\it Z. Physik }{\bf C #1} (19#2) #3}
\def\mpla#1#2#3{   {\it Mod. Phys. Lett. }{\bf A #1} (19#2) #3}

\def\ibid#1#2#3{   {\it ibid. }{\bf #1} (19#2) #3}
\def\ijmpa#1#2#3{  {\it Int. J. Mod. Phys. }{\bf A #1} (19#2) #3}
\def\ltap{\ \raisebox{-.4ex}{\rlap{$\sim$}} \raisebox{.4ex}{$<$}\ }
\def\gtap{\ \raisebox{-.4ex}{\rlap{$\sim$}} \raisebox{.4ex}{$>$}\ }
\def\eq#1{{eq.~(\ref{#1})}}
\def\eqs#1#2{{eqs.~(\ref{#1})--(\ref{#2})}}

\let\vev\VEV

\let\mod\abs
\def\Im{\mathop{\mbox{Im}}}
\def\Re{\mathop{\mbox{Re}}}
\def\Tr{\mathop{\mbox{Tr}}\,}
\def\etal{{\it et al.}}

\newcommand{\bea}{\begin{eqnarray}}
\newcommand{\beq}{\begin{equation}}
\newcommand{\eea}{\end{eqnarray}}
\newcommand{\eeq}{\end{equation}}
\newcommand{\nnu}{\nonumber}
\newcommand{\spav}[1]{\parbox{1mm}{\vspace*{#1}}}

\begin{document}

\begin{titlepage}
\begin{flushright}
{\tt CERN-TH.7306/94 \\
SISSA 83/94/EP}
\end{flushright}
\spav{0.8cm}
\begin{center}
{\Large\bf Studying $\varepsilon '/\varepsilon$ in the Chiral Quark Model:}\\
\spav{0.5cm}
{\large\bf $\gamma_5$-Scheme Independence and NLO Hadronic Matrix Elements}\\
\spav{1.5cm}\\
 {\Large Stefano Bertolini}
\spav{1cm}\\
{\em INFN, Sezione di Trieste, and}\\
{\em Scuola Internazionale Superiore di Studi Avanzati}\\
{\em Via Beirut 4, I-34013 Trieste, Italy}\\
\spav{.5cm}\\
{\Large Jan O. Eeg\fnote{\dag}{On leave of absence from the Dept. of Physics,
University of Oslo, N-0316 Oslo, Norway.} and Marco Fabbrichesi}
\spav{1cm}\\
{\em CERN, Theory Division}\\
{\em CH-1211 Geneva 23, Switzerland}\\
\spav{1.9cm}\\
{\sc Abstract}
\end{center}

We study the $CP$-violating, $|\Delta S| =1$ parameter
$\varepsilon '/\varepsilon$ by computing the
hadronic matrix elements in the chiral quark model.
We estimate in the chiral expansion   the
coefficients of  the next-to-leading order terms
that correspond to the operators
$Q_6$ and $Q_{11}$. We consider the impact of these corrections on the
value of $\varepsilon '/\varepsilon$.
We also investigate the possibility that the chiral quark model
might drastically reduce the
dimensional regularization $\gamma_5$-scheme dependence of
current evaluations of $\varepsilon '/\varepsilon$.

 \vfill
\spav{.4cm}\\
{\tt CERN-TH.7306/94\\
SISSA 83/94/EP \\
August 1994}

\end{titlepage}

\newpage
\setcounter{footnote}{0}
\setcounter{page}{1}

\section{Foreword and Summary}

A detailed study of the
$CP$-violating, $|\Delta S| =1$ parameter $\varepsilon '/\varepsilon$
puts us at the crossroad of different aspects of
standard model physics
and, accordingly, provides a promising
 testing ground of model-dependent features.

As the precision in the experimental measurements of $\varepsilon
'/\varepsilon$ improves~\cite{eps}, all theoretical predictions face the
challenge of reducing their uncertainties. These have a twofold
 origin as,
on the one
hand, the  analysis of the Wilson coefficients must be pushed to
higher orders in $\alpha_s$ and $\alpha$
 (as well as
including all the relevant operators) while,
on the other hand, the estimate of the hadronic matrix
elements has to be substantially improved.

Indeed, a great deal of work has been done in this direction. In
recent years
the Wilson coefficients of the effective
lagrangian for $|\Delta S| = 1$
weak decays have been computed to the next-to-leading order
(NLO)~\cite{Monaco,Roma},
and matrix
elements have been calculated by a variety of techniques
 like $1/N_c$~\cite{1/N}, quark
models~\cite{PdeR91} and lattice~\cite{Roma}. Yet it is perhaps fair to say
that, all efforts notwithstanding, the current
uncertainty in the theoretical prediction of $\varepsilon '/\varepsilon$ is
still large, most of it arising from the hadronic matrix elements.

There are at least four relevant
perturbative expansions which are important in estimating
hadronic matrix elements. These are the
$1/N_c$ and $\alpha_s/\pi$ expansions  for non-factorizable contributions,
 which have been studied in~\cite{1/N} and \cite{PdeR91},
respectively. Then we have the chiral  expansion,
which comes in two varieties
according to the energy scale at which it is considered. The
coefficients of the chiral lagrangian take contributions from chiral
loops (corresponding to momenta below the scale $\Lambda_{QCD}$ at
which all quark
fields are integrated out) as well as
from light-quark loops (between $\Lambda_{QCD}$
and the chiral symmetry breaking scale
$\Lambda_\chi$).

The chiral quark model ($\chi$QM)~\cite{leqcd,ERT,BBdR}
can be thought as interpolating between the
lower-end of the perturbative short-distance analysis ($\approx 1$ GeV)
and the chiral lagrangian energy region ($\Lambda_{QCD}\approx M$).
In the model, the scale $M$
is identified with the constituent quark mass. This is
the only free parameter in our analysis and most
matrix elements are a sensitive function of it.
What are the allowed values for this parameter?
As we shall discuss in section 2, a rather large range
\mbox{($100 < M < 400$ MeV)}
is possible,
different analyses having different outcomes according to
 the actual physical processes considered.

In this paper, we begin a study of
$\varepsilon'/\varepsilon$ in the $\chi$QM by focusing on two aspects: the
relevance of the NLO corrections to the hadronic matrix elements and the
problem of the $\gamma_5$-scheme independence of the estimate.
Our original motivation was to study
the gluonic magnetic dipole penguin operator
 \beq
Q_{11} \simeq \frac{g_s}{16\pi^2} \, m_s \: \overline{s}\, \sigma_{\mu\nu}
t^a G^{\mu\nu}_a (1-\gamma_5 )\, d \label{Q11} \, ,
\eeq
which until recently has been tacitly neglected in all discussions of
$\varepsilon '/\varepsilon$.
The potential relevance of the
operator $Q_{11}$  for $\varepsilon '/\varepsilon$  has been
discussed in a previous work~\cite{BFG}. We confirm the result of
ref. \cite{BFG} up to an overall sign.
In section 3 we discuss in detail the bosonization of $Q_{11}$.
The
first non-vanishing contribution of $Q_{11}$ to the $K\to \pi\pi$ matrix
elements comes from a term that is NLO in the chiral expansion (O$(p^4)$).
Even though the corresponding coefficient
of the chiral expansion---as well as the short-distance
Wilson coefficient---is large, the matrix element turns out to be
kinematically suppressed by a factor $m_\pi^2/m_K^2$ with respect to the
dominant operators, making its contribution to $\varepsilon '/\varepsilon$
subleading.

In order to better gauge the significance of this NLO contribution,
 we  estimate, in section 4, the
NLO correction to the matrix element of the ordinary gluonic penguin
operator
\beq
Q_6 =  \overline{s}_\alpha \gamma_\mu ( 1 - \gamma_5 ) d_\beta \sum_q
\overline{q}_\beta
\gamma^\mu ( 1 + \gamma_5 ) q_\alpha \, ,\label{Q6}
\eeq
which gives the dominant contribution to  $\varepsilon '/\varepsilon$.
Such an NLO correction to the hadronic matrix element of $Q_6$ turns out to
be generally larger
than the contribution of the operator $Q_{11}$, as well
as the leading-order (LO) contributions of the
operators $Q_3$, $Q_5$ and $Q_7$ (see section 5
for definitions). Its size increases monotonically with $M$,
ranging from 14\% of the LO term
for $M=120$ MeV up to 25\% and more for
$M \gtap 220$ MeV. Our perturbative results (we perform an expansion in
powers of $M^2/\Lambda_\chi^2$)
should not be trusted for  larger values of $M$,
higher-order corrections being needed.

Our approach makes it also possible to study the dimensional regularization
$\gamma_5$-scheme
dependence of the matrix elements of these operators. This is an important
problem since it causes, if we take the NLO results for the Wilson
coefficients in the two schemes,
 a significant part of the current
theoretical uncertainty (see for instance ref. \cite{Monaco})---as large
as 70--80\% for $m_t = 170$ GeV. One expects this scheme dependence to
disappear
when matrix elements are consistently evaluated~\cite{NDR}.
This cannot be done within
the lattice or $1/N_c$ approach.

As a preliminary analysis, in section 5
we use the $\chi$QM to study the  $\gamma_5$-scheme
dependence of $\varepsilon'/\varepsilon$ in a toy model, made of
the LO matrix elements of the operators $Q_6$ (for the $I=0$ amplitude)  and
of the electroweak penguin operator (for $I=2$):
\beq
Q_{8} = \frac{3}{2} \, \overline{s}_{\alpha}
\gamma_\mu ( 1 - \gamma_5 ) d_{\beta}
         \sum_{q} e_q\ \overline{q}_{\beta}
\gamma^\mu ( 1 + \gamma_5 ) q_{\alpha}
\, ,\label{Q8}
\eeq
which becomes important for a heavy top quark.
As a matter of fact, the Wilson coefficients of these two operators exhibit
the largest scheme dependence. In comparison, the scheme dependence of the
Wilson coefficients of $Q_4$ and $Q_9$,
next in relevance to $Q_6$ and $Q_8$,
is ten times smaller.

We find that the
$\gamma_5$-scheme dependence of $\varepsilon '/\varepsilon$
in our toy model is
reduced below the 10\% level for values of the
parameter $M$ well within the allowed range. This means that,
for any value of
$\Lambda_{QCD}^{(4)}$ in the 200--400 MeV range and of the
renormalization scale $\mu$ between 1 GeV and $m_c$,
it is possible to find a value of $M$ in a restricted
range (120--220 MeV)
for which the scheme dependence of the hadronic matrix elements
exactly cancels that of the Wilson coefficients.

Notice that in the $\chi$QM
the matrix element of $Q_{11}$ turns out to be scheme-in\-de\-pen\-dent,
as it happens
for its ``LO'' Wilson coefficient~\cite{C11schemeind}.

The range of values for which we find
scheme independence is consistent with
the values of $M$ for which we can trust our
perturbative results at the order $M^2/\Lambda_\chi^2$
\mbox{($M\ltap 200$ MeV).}
To extend our analysis in a reliable manner to  values of $M$ of
300 MeV and more requires considering higher-loop
effects---a difficult computational problem.

As encouraging as these results might be, we should remember that
we are considering a toy model for $\varepsilon '/\varepsilon$
in which only the two
leading operators $Q_6$ and $Q_8$  are included.
In order to ascertain conclusively the scheme independence and the size
of $\varepsilon '/\varepsilon$, all relevant operators should be evaluated in
the $\chi$QM and consistently included.
Preliminary estimates show that such an extension
stabilizes the range of values of $M$ for which the scheme
independence is substantially weakened, further reducing the renormalization
scale dependence.

In the final part of the paper (section 6)
we compare the contribution
to $\varepsilon '/\varepsilon$ of the dipole penguin operator
$Q_{11}$ to the other ten traditional operators.
Since the hadronic matrix elements of the other operators
beside $Q_6$ are known in the $\chi$QM
only to the leading factorizable order,
we follow for the ten traditional operators the $1/N_c$ approach
of ref.~\cite{Monaco}.

A detailed anatomy of all individual contributions is summarized
in tables 6, 7 and 8. We find these tables a useful way of displaying the
available information and a standard reference for future developments.

The Wilson coefficients are computed to the NLO level, at least to the extent
made possible by the entries of the anomalous dimensions currently
available---the three-loop
NLO mixing between $Q_{11}$ and the other operators and related two-loop
initial conditions are still missing.
This may amount to a 40--50\% uncertainty in the
$Q_{11}$ Wilson coefficient at the 1 GeV scale (extrapolating the
results for $b\to s\gamma$ given in ref.~\cite{nlobsgamma}).
Nonetheless, this little affects  the
final result for $\varepsilon '/\varepsilon$ since the evolution of the
first ten Wilson coefficients is independent on the presence of the
$Q_{11}$ (and $Q_{12}$) operator.

 Our analysis shows that the
inclusion of the magnetic penguin operator affects the prediction of
$\varepsilon '/\varepsilon$ always below the
$10^{-4}$ level, therefore putting $Q_{11}$ in
the same class as other operators like  $Q_3$, $Q_5$ and $Q_7$, whose
contributions are even smaller;
yet, for $m_t \gtap$ 170 GeV, the
size of $\varepsilon '/\varepsilon$ itself becomes of the same order as the
$Q_{11}$ contribution.

The completion of the
study of $\varepsilon '/\varepsilon$ in the $\chi$QM approach
is needed to ascertain conclusively the $\gamma_5$-scheme independence
of the result and, what is equally relevant, the corresponding size
of $\varepsilon '/\varepsilon$. Corrections of the order of the current quark
masses and $\alpha/\pi$ will then have to be included. We expect the mass
corrections to be of the order or less than 10\%, while those of
(soft) gluon contributions to be larger ($\ltap 50$\%).
We recall that in our toy model we find
systematically a factor of 3--4 enhancement with respect
to the average of the $1/N_c$
results with $Q_6$ and $Q_8$ only and $m_t=170$ GeV. This is
partially
due to a reduced cancellation between the two operators in the $\chi$QM.

We also stress the relevance of the calculation of NLO corrections
to all relevant matrix elements, which become crucial for $m_t\gtap 170$
GeV, due to the large cancellations that occur between gluonic and
electromagnetic penguin contributions.

\section{Bosonization of quark operators}

In general, the effective lagrangian at the quark level has the form

\begin{equation}
{\cal{L}}_{Q} \, = \,  \sum_i \, C_i \; Q_i  \; ,
\end{equation}
where the coefficients $C_i$ contain the short-distance physics above the
renormalization point $\mu \simeq 1$ GeV, and $Q_i$ are operators containing
quark fields (at the scale $\mu$).
When  the quark operators are bosonized, they
will be
 represented by some linear combination of meson operators
${\cal{L}}_j^{(\chi)}$:
\begin{equation}
Q_i \, \rightarrow  \,  \sum_j \, b_{ij} \; {\cal{L}}_j^{(\chi)} \, ,
\end{equation}
leading to the following chiral lagrangian at the meson level:
\begin{equation}
{\cal{L}}_{Q} \, \rightarrow  \, {\cal{L}}_{\chi \mbox{\scriptsize PT}} \, = \,
 \sum_{i,j} \, G_j (Q_i) \, {\cal{L}}_j^{(\chi)} \, = \,
 \sum_j \, G_j \, {\cal{L}}_j^{(\chi)} \, ,
\eeq
where $G_j(Q_i)  \equiv  C_i \, b_{ij}$ (no sum on $i$).

One way to perform the bosonization is to
 give the complete list of operators
${\cal{L}}_j^{(\chi)}$ to  which some quark operator
$Q_i$ could a-priori contribute \cite{enp,bep,cl}.
 To determine the various coefficients $b_{ij}$, one  tries combinations
of experimental data and  phenomenological estimates.
In this work, we have chosen a version of the $\chi$QM
 advocated by various authors~\cite{leqcd,ERT}.
This type
of effective low-energy model
 is obtained by adding  a new term to the ordinary QCD lagrangian:
\begin{equation}
  {\cal{L}}_{\chi \mbox{\scriptsize QM}} = - M \left( \overline{q}_R \; \Sigma
q_L +
\overline{q}_L \; \Sigma^{\dagger} q_R \right) \; , \label{4}
\end{equation}
where $\overline{q} = (\overline{u},\overline{d},\overline{s})$,
 and the $3\times 3$ matrix
$\Sigma \equiv \exp \left( 2i\, \Pi /f \right)$
contains the pseudoscalar octet mesons
 $\Pi = \sum_a \pi^a\lambda^a/2 $
$(a=1,..,8)$. The scale
$f$ is identified with the (bare) pion decay constant,
so that numerically $f \simeq   f_\pi \simeq 93.3$ MeV
(and equal to $f_K$ in the chiral limit).

The parameter $M$ is characteristic of the $\chi$QM and
represents a typical constituent quark mass for the light quarks.
For $M\neq 0$ the chiral symmetry is broken and the quark condensate acquires a
non-vanishing vacuum expectation value.
A central value of 320 MeV is found
in ref.~\cite{PdeR91}, albeit with a comparable uncertainty. Smaller values
are found by
fitting data relevant to meson physics,
as in ref.~\cite{Bijn}, where \mbox{$M = 223 \pm 9$ MeV} is quoted.
It seems fair  to consider the range
$100\,\mbox{MeV} < M < 400\,\mbox{MeV}$, as a reasonable present-day
uncertainty on the determination of the parameter $M$.

The resulting field theory with
pseudoscalar quark-meson couplings (see appen\-dix~A)
and quark loops gives a good description
of the  $K \rightarrow \pi \pi$ amplitudes to leading order in $1/N_c$
\cite{PdeR91}, and under certain conditions of the anomalous
 $\pi^0 \rightarrow \gamma \gamma$ amplitude \cite{anomaly,Bijn},
as well as $K^0 \rightarrow \gamma\gamma$~\cite{ep93,ep94a}.
It is also worth recalling that the $\Delta = 1/2$ rule in $CP$-conserving
kaon decays is reproduced within the model up to factors 2--3~\cite{PdeR91},
whereas other approaches fail by an order of magnitude.

When performing loop integrals within this model,
the logarithmic and quadratic
divergences are identified with $f_{\pi}$ and the quark condensate
$\langle \overline{q}q \rangle$, respectively, that is:
\begin{equation}
f_{\pi}^{(0)} \, = \, \frac{N_c M^2}{4 \pi^2 f} \hat{f}_{\pi}
 \qquad \mbox{and} \qquad
\vev{\overline{q} q} ^{(0)} \, = \, \frac{N_c M}{4 \pi^2} \hat{C}_q \; ,
\label{fqq}
\end{equation}
where $f^{(0)}_\pi$ and  $\langle \overline{q}q \rangle ^{(0)}$
 are the pion decay constant and the quark condensate to zeroth order
in $\alpha_s$ (no gluon-condensate corrections and
zero current-quark masses, see \eq{qqmass}).
We distinguish between $f^{(0)}_{\pi}$ and $f$
because of their different origin, even though at out level of
approximation $f^{(0)}_{\pi} = f \simeq f_{\pi}$.

In dimensional regularization ($d=4-2\epsilon$)
we have
\beq
\hat{f}_{\pi}= \frac{1}{\epsilon} - \gamma_E +\ln 4 \pi + \ln
\frac{\mu^2}{M^2} \qquad \mbox{and} \qquad
\hat{C}_q = M^2(\hat{f}_{\pi}+1) \, . \label{dimreg}
\eeq
 It is perhaps useful to recall that, introducing in the effective theory
a cut-off $\Lambda$ one obtains~\cite{BBdR}
\beq
\hat{f}_{\pi}= \log (\Lambda^2/M^2) + ... \qquad \mbox{and} \qquad
\hat{C}_q = - \Lambda^2 + M^2 \log (\Lambda^2/M^2)  + ...  \label{cutoff}
\eeq
To the order we work we cannot remove the cut-off. Equations (\ref{fqq}) are
thus a convenient book-keeping device to identify $f_\pi$
and $\vev{\overline{q}q}$ in the loop integrals.

The model has a  ``rotated''  picture, where
 the term ${\cal{L}}_{\chi \mbox{\scriptsize QM}}$
in (\ref{4}) is transformed into a pure
mass term $- M \overline{{\cal{Q}}} {\cal{Q}}$
for rotated "constituent quark" fields ${\cal{Q}}_{L,R}$ :
\beq
q_L \rightarrow  {\cal{Q}}_L =  \xi  q_L \quad \mbox{and} \quad
q_R \rightarrow  {\cal{Q}}_R =  \xi^\dagger q_R \, ,
\eeq
where $\xi \, \cdot \, \xi  = \Sigma $.
The meson--quark couplings in this rotated  picture  arise from
the kinetic
(Dirac) part of the constituent quark lagrangian.
These interactions can be described in terms of vector and axial vector
fields coupled to constituent quark fields
${\cal{Q}} = {\cal{Q}}_R + {\cal{Q}}_L$:
\beq
{\cal{L}}_{int} \, = \, \overline{{\cal{Q}}} \Bigl[ \gamma^{\mu}
{\cal{V}}_{\mu}
\,  + \,
\gamma ^{\mu} \gamma _5   {\cal{A}}_{\mu} \, \Bigr] {\cal{Q}} \, , \label{7}
\eeq
where
\beq
{\cal{V}}_{\mu} \, = \, (R_{\mu} + L_{\mu})/2  \, , \qquad
{\cal{A}}_{\mu} \, = \, (R_{\mu} - L_{\mu})/2  \, ,
\eeq
and
\beq
L_{\mu} \, = \, \xi \, (i \partial_{\mu} \, \xi^\dagger) +
 \xi \, l_\mu \, \xi^\dagger \, , \qquad
R_{\mu} \, = \, \xi^\dagger (i \partial_{\mu} \xi) + \xi^\dagger
r_\mu \, \xi \, .
\eeq
Here $l_\mu$ and $r_\mu$ are the external fields containing the photon
(as well as the $W$ field).

Using (\ref{7}), the  strong chiral lagrangian  $O (p^2)$
 can be understood
as two axial currents ${\cal{A}}_{\mu}$ attached to a quark loop, leading
to
\begin{equation}
{\cal{L}}^{(2)}_{s} \, \sim \,  \Tr \Bigl[ {\cal{A}}_{\mu} \, {\cal{A}}^{\mu}
 \Bigr] \, .
\end{equation}
Using the relations \cite{Bijn,brupr}
\begin{equation}
   2 i \, {\cal{A}}_{\mu} \, = \; - \,
\xi^{\dagger} (D^{\mu} \Sigma) \xi^{\dagger} \; = \;
 \xi (D_{\mu} \Sigma^{\dagger}) \xi \, , \label{9}
\end{equation}
one obtains the leading strong chiral lagrangian
\begin{equation}
{\cal{L}}^{(2)}_{s} \, = \, \frac{f^2}{4} \,
 \Tr \Bigl( D^{\mu} \Sigma^{\dagger} \, D_{\mu} \Sigma \Bigr) \; ,
\end{equation}
where $D_\mu$ is the covariant derivative.
Note  that
${\cal{A}}_\mu$ is invariant under
local chiral transformations~\cite{leqcd},
 in agreement with the invariance of
 ${\cal{L}}^{(2)}_s$. In contrast,
 the vector field ${\cal{V}}_\mu$ transforms as a gauge field.
Therefore, terms involving the vector field would manifestly
break  local chiral invariance and cannot appear in a chiral lagrangian,
except for the anomaly term~\cite{ep94a}.
 The vector field can only be present indirectly
through some field tensor obtained from the commutator of the total
covariant derivative of the  lagrangian.
 In addition, the covariant derivative of
${\cal{A}}_\mu$ may occur in higher-order
terms~\cite{PdeR91,cl,brupr,DS}.

 In the version of the $\chi$QM given by (\ref{4})
(the unrotated  picture)
the momenta which correspond to derivatives of the fields have to be
extracted from the quark propagators in the loop diagram,
 and it is in general difficult to see the correspondence between the
quark and the bosonized version of the operators.
The rotated picture, where the
axial vector fields in \eq{7} couple
to the quark loops with derivative couplings, is more
transparent in this sense. For example, for
$O (p^4)$ terms (with two momenta
and one current quark mass), no momenta have to be extracted from the
quark propagators, and
we can readily deduce the chiral lagrangian (bosonized quark operator)
in terms of the rotated fields.

\section{Bosonization of $Q_{11}$}

The magnetic dipole operator (\ref{Q11}) can be rewritten as
\begin{equation}
 Q_{11} \, = \, \frac{g_s}{8 \pi^2}
 \overline{s} \; \Bigl[ m_d \, R + m_s \, L \Bigr]
 \,  \, \sigma  \cdot  G
 \;  d  \, + \, h.c. \; , \label{11}
\eeq
where $ \sigma  \cdot \, G  \equiv
\sigma^{\mu \nu}\, G^a_{\mu \nu} t^a$,
 $G^a_{\mu \nu}$ is the gluon field, $t^a$ are the
SU(3) generators, normalized as $\Tr ( t^a t^b ) = (1/2) \delta^{ab}$,
 and $(L,R)=(1\mp\gamma_5)/2$.
The  Hermitian conjugate is there to remind us of the
corresponding $s \rightarrow d$ transition. This will be understood
throughout the paper.
The dipole  operator in \eq{11} can be written in a
chiral  $SU(3)$
covariant form as
\begin{equation}
 Q_{11} \, = \,   \frac{g_s}{8 \pi^2} \, \Bigl[
 \overline{q}_R \; {\cal{M}}_{q} \lambda_-
 \; \sigma  \cdot  G \, q_L \, + \,
\overline{q}_L \;
 \; \sigma  \cdot  G \;
\lambda_- {\cal{M}}_{q}^{\dagger}  \;  q_R \Bigr] \, , \label{12}
\end{equation}
where  ${\cal{M}}_{q} \, = \,\mbox{diag} \,(m_u,m_d,m_s)$ is the current
quark matrix.
The Gell-Mann matrices
$\lambda_{\pm} \, = \, (\lambda_6 \pm i \, \lambda_7)/2$ project
$\Delta S \, = \, \pm 1$ transitions  out of the quark fields
 $\overline{q}_L \, = \,
(\overline{u}_L, \, \overline{d}_L, \, \overline{s}_L )$.
 Note that in \eq{12} the
 operator transforms as $(\underline{8}_L, \underline{1}_R)$ under
the chiral $SU(3)_L \times SU(3)_R$ symmetry if
the current quark matrix is taken to transform as
${\cal{M}}_{q}\to U_R{\cal{M}}_{q}U_L^\dagger$, where $U_R$ and $U_L$ are the
chiral $SU(3)$ transformation matrices.

In the rotated picture, $Q_{11}$ reads
\beq
Q_{11} \, = \, \frac{g_s}{8 \pi^2} \, \overline{{\cal{Q}}} \Bigl[ F_{(-)}^V \,
+
 \, F_{(-)}^A \, \gamma_5 \Bigr] \:
\sigma  \cdot  G \;
 {\cal{Q}} \, ,
\eeq
where $F_{(-)}^{V,A} = \left(F_{(-)}^R \pm F_{(-)}^L\right) /2$, with
\beq
 F_{(-)}^L \; = \,  \xi^{\dagger} \, {\cal{M}}_{q}
         \lambda_- \xi^{\dagger}  \, , \qquad \mbox{and} \qquad
 F_{(-)}^R \; = \; \xi \, \lambda_-
{\cal{M}}_{q}^{\dagger} \,  \xi \, . \label{13}
\eeq

The operator corresponding to $Q_{11}$
is understood in terms of a quark loop.
Let us first note that the lowest-order
contribution is given by a quark loop where only the field
$F_{(-)}^V$ is interacting
\begin{equation}
 {\cal{L}}^{(2)}(Q_{11}) \, \sim \, \Tr (F_{(-)}^V)  \; = \;
 \Tr \Bigl[ \Sigma^{\dagger} \, {\cal{M}}_{q} \, \lambda_- +  \lambda_- \,
\label{14}
{\cal{M}}_{q}^{\dagger} \,  \Sigma \Bigr] \; .
 \end{equation}
This is the contribution that cancels against the pole term induced
by a non-vanishing $\vev{0|Q_{11}|K^0}$ transition~\cite{Don Hol},
in agreement with the FKW theorem~\cite{FKW}.
To the NLO, we obtain a term corresponding to an interaction
of $F_{(-)}^V$ and two axial fields attached to a quark loop:
\begin{equation}
 {\cal{L}}^{(4)}(Q_{11}) \, \sim \,
\Tr \Bigl[ F_{(-)}^V  {\cal{A}}_{\mu} \, {\cal{A}}^{\mu} \Bigr]
  \,  . \label{15}
\end{equation}
Note that $F_{(-)}^A$ is not contributing in
\eqs{14}{15}, because there must be an even number of
 $\gamma_5$'s in the quark loop.
 Using (\ref{9}) and (\ref{13}), we find that
${\cal{L}}^{(4)}(Q_{11})$ can be written in the form
\begin{equation}
 {\cal{L}}^{(4)}(Q_{11})    =
G_{\underline{8}}^{(4)} (Q_{11}) \, \Tr  \Bigl[\left(
\Sigma^{\dagger} \, {\cal{M}}_{q} \lambda_-
                 +   \lambda_- {\cal{M}}_{q}^{\dagger}\Sigma\right)
  \, D^{\mu} \Sigma^{\dagger} D_{\mu} \Sigma  \Bigr]  \, .
\label{16}
\end{equation}

In \eq{12}, we see that $\lambda_-$ and ${\cal{M}}_{q}$ are always
next to each other, and so they must be in the bosonized version
of $Q_{11}$.
 We  neglect the
possible terms appearing as the product of two traces. The latter
correspond to two quark loops connected by (soft) gluons and will therefore be
suppressed by a factor
$(\alpha_s/\pi)^2$, at least.
Due to the $CPS$ symmetry~\cite{CPS} of the quark operators,
also the corresponding bosonized operators have
to exhibit such a symmetry. As a consequence the
two terms in \eq{16} must appear with equal weight.
 Among all the  possible terms
for $|\Delta S| = 1$ kaon decays, which can be obtained by
inserting the flavor-changing  matrix $\lambda_-$
into various places within the trace in
${\cal{L}}^{(4)}_{s}$ \cite{cl}, we are thus left with the unique term in
\eq{16}.

The coefficient in \eq{16}  can be most easily calculated in the
unrotated picture, by considering, for instance, the off-shell $K
\rightarrow \pi$
transition.
In order to calculate  $ G_{\underline{8}}^{(4)}(Q_{11})$,
we expand  ${\cal{L}}^{(4)}(Q_{11})$
to find the amplitude
\begin{equation}
{\cal A}(K^+ \rightarrow \pi^+ ; Q_{11}) \, = \,
\vev{\pi^+|i\ {\cal L}_\chi(Q_{11})|K^+} \, = \,
\frac{2 i}{f^2} \, (m_s +m_d) \, k^2 \,
G_{\underline{8}}^{(4)} (Q_{11}) \, , \label{3.8}
\end{equation}
where $k$ is the off-shell momentum.
Matching \eq{3.8} with the corresponding quark loop amplitude, represented by
the diagrams in Fig. 1, leads to
\begin{equation}
G_{\underline{8}}^{(4)} (Q_{11}) \, = \, - \, \frac{11}{4} \,
\langle \overline{q} q\rangle_G \, \left(
\frac{C_{11}}{16\pi^2} \right) \, ,
\end{equation}
where we have used
\begin{equation}
\langle \overline{q} q\rangle_G
\; \;  \equiv \; \;
 - \frac{1}{12M} \, \langle \frac{\alpha_s}{\pi}G G \rangle \ ,
\label{GGqq}
\end{equation}
which represents the two-gluon condensate contribution to the
quark condensate (second diagram in Fig. 2).
Diagrams analogous to those in Fig. 1
with the two meson lines entering at the same point
do not contribute to momentum dependent terms, and are disregarded.
In Table 1 we show the values of $\langle \overline{q} q\rangle_G$
as a function of $M$. For the gluon condensate we have taken the
central value of the lattice evaluation
$\langle \frac{\alpha_s}{\pi}G G \rangle = (460\pm 20)^4$
MeV$^4$~\cite{GGlattice}. The entries shown can be  scaled accordingly
for other values of the gluon condensate.

Having determined $G_{\underline{8}}^{(4)} (Q_{11})$ from the
 $K \rightarrow \pi$ transition,
we can deduce the $K \rightarrow 2 \pi$ amplitude from the chiral structure
of ${\cal{L}}^{(4)}(Q_{11})$:
\begin{equation}
{\cal A}(K^0 \rightarrow \pi^+ \pi^- ; Q_{11}) \, = \,
\frac{\sqrt{2}}{f^3} \, (m_s - m_d)
 \, m_{\pi}^2 \,\, G_{\underline{8}}^{(4)} (Q_{11}) \; , \label{2.13}
\end{equation}

Even though the coefficient in ${\cal{L}}^{(4)}(Q_{11})$ is large,
the $Q_{11}$ contribution to $K \rightarrow 2\pi$ is small because of the
factor $m_{\pi}^2$ (in place of $m_K^2$ obtained for $Q_6$).
The modest role played
by $Q_{11}$ in $\varepsilon '/\varepsilon$
comes therefore from this kinematical
suppression rather than from its contribution being NLO in the chiral
expansion.

\section{Next-to-leading order bosonization of $Q_{6}$}

The bosonization of $Q_6$ follows basically the same line as for
$Q_{11}$ in the preceding section. The standard expression (obtained from
\eq{Q6} by a Fierz transformation)
 \begin{equation}
 Q_6 \, = \,  - \, 8 \, \sum_q
 (\overline{s}_L q_R) \, (\overline{q}_R d_L)  \; ,
\end{equation}
 can be rewritten  in the rotated picture as
\beq
Q_6 \, = \, - \, 8 \, (F_{(-)})_{\alpha \beta}
 \, (\overline{\cal{Q}}_L)_{\alpha} ({\cal{Q}}_R)_{\delta}
(\overline{\cal{Q}}_R)_{\delta} ({\cal{Q}}_L)_{\beta} \, ,
\eeq
where $F_{(-)} \; = \,  \xi \, \lambda_- \xi^{\dagger}$,
and the greek letters are flavor indices.
 Thus, likewise to ${\cal{L}}^{(4)}(Q_{11})$, the chiral representation
of $Q_6$ to leading order can be written as
\begin{equation}
 {\cal{L}}^{(2)}(Q_{6}) \, \sim \,
\Tr \Bigl [F_{(-)}  {\cal{A}}_{\mu} \, {\cal{A}}^{\mu} \Bigr]  \, , \label{23}
\end{equation}
which by means of \eq{9} can be written in the same
familiar form as the other
$|\Delta S = 1|$  octet operators  $O (p^2)$~\cite{cro67}:
\begin{equation}
{\cal{L}}^{(2)}_{\Delta S = 1} \, = \, G^{(2)}_{\underline{8}}(Q_6) \,
 \Tr \Bigl( \lambda_- D^{\mu} \Sigma^{\dagger} D_{\mu}
 \Sigma \Bigr) \, ,
\label{L2(Q6)}
\end{equation}
where $ G^{(2)}_{\underline{8}} (Q_6)$ is the most important component.
The term in \eq{L2(Q6)} gives rise to the amplitudes
\bea
{\cal A}( K^+ \rightarrow \pi^+ ;\, Q_6) & = &
\frac{2i}{f^2}\ k^2\ G_{\underline{8}}^{(2)}(Q_6) \, ,
\label{A2(Q6)o} \\
{\cal A}( K^0 \rightarrow \pi^+ \pi^- ;\, Q_6) & = &
\frac{\sqrt{2}}{f^3}  \Bigl[ m_K^2 - m_\pi^2
\Bigr]\, G_{\underline{8}}^{(2)}(Q_6) \, .
\label{A2(Q6)}
\eea

Now, we want to find the bosonization of $Q_6$ to the same order as
${\cal{L}}^{(4)}(Q_{11})$ in the preceding section. That is,
the bosonized operator has to contain a mass insertion in addition to what is
already included in ${\cal{L}}^{(2)}(Q_{6})$. The QCD mass lagrangian can be
written as
\begin{equation}
 {\cal{L}}_{\ \mbox{\scriptsize mass}} \, = \, - \,
\left[ \overline{q}_R \, {\cal{M}}_{q}  \, q_L \, + \,
\overline{q}_L \, {\cal{M}}_{q}^{\dagger} \, q_R \right]  \; , \label{25}
\end{equation}
which can be transformed to the form
${\cal{L}}_{\ \mbox{\scriptsize mass}} \, = \, - \, \overline{{\cal{Q}}}
 \widetilde M_q  {\cal{Q}}$,
where
\beq
 \widetilde M_q   \; \equiv  \; \xi^{\dagger} \, {\cal{M}}_{q}
  \xi^{\dagger} \, L  \;   +  \;   \xi \,
{\cal{M}}_{q}^{\dagger} \,  \xi \, R \, .
\eeq
 Therefore, a possible NLO representation of $Q_6$ is given by
\begin{equation}
 {\cal{L}}^{(4)}(Q_{6}) \, \sim \,
\Tr  \Bigl[ F_{(-)} \,  {\cal{A}}_{\mu} \, \widetilde M_q^X \, {\cal{A}}^{\mu}
\Bigr]  \,  ,
\label{27}
\end{equation}
with the addition of two other terms, where the quantities within the
 trace are permuted. In \eq{27}
 $\widetilde M_q^X$ represents some part of $\widetilde{M_q}$ (i.e.
$X=R,L,V,A$).
Using \eq{9}, we obtain three different $CPS$-symmetric terms:
\begin{equation}
 {\cal{L}}^{(4)}_E  (Q_6) \; = \;
G_E^{(4)}(Q_6) \,
\Tr \Bigl[\left(\Sigma^{\dagger} \, {\cal{M}}_{q} \lambda_-  \,  +
  \; \lambda_- {\cal{M}}_{q}^{\dagger} \,\Sigma \right)  \,
D^{\mu} \Sigma^{\dagger} D_{\mu} \Sigma \Bigr]  \; ,
\end{equation}
\begin{equation}
 {\cal{L}}^{(4)}_H (Q_6)  \; = \;
G_H^{(4)}(Q_6) \, \Tr \Bigl[\left( \lambda_- \, \Sigma^{\dagger} \,
{\cal{M}}_{q} \,  +
 \;  {\cal{M}}_{q}^{\dagger} \,\Sigma \, \lambda_- \right) \,
D^{\mu} \Sigma^{\dagger} D_{\mu} \Sigma \Bigr]  \; ,
\end{equation}
\begin{equation}
 {\cal{L}}^{(4)}_K (Q_6)   \; = \;
G_K^{(4)}(Q_6) \, \Tr \Bigl[\left(\Sigma \, {\cal{M}}_{q}^{\dagger} \;
+  \; {\cal{M}}_{q} \,\Sigma^{\dagger} \right)  \,
D^{\mu} \Sigma\, \lambda_-\, D_{\mu} \Sigma^{\dagger} \Bigl]  \, . \label{LK}
\end{equation}

 These three terms are in fact linear combinations
of those given in ref. \cite{cl}
(again we have discarded subleading terms which are the product of two traces).
Note that the last term is not immediately obtained by inserting $\lambda_-$
within the trace in ${\cal{L}}^{(4)}_s$; in order to obtain (\ref{LK}),
 one has to use
$ D^{\mu} \Sigma \, = \, -
\Sigma \, (D_{\mu} \Sigma^{\dagger}) \, \Sigma$ and
$ 1 = \Sigma \, \Sigma^{\dagger} \rightarrow
\Sigma \, \lambda_- \, \Sigma^{\dagger}$.
In the case of ${\cal{L}}^{(4)} (Q_{11})$ it was sufficient to calculate
the  $K \rightarrow \pi$ transition induced by $Q_{11}$  to
determine the unique coefficient. Here we have to find
 three coefficients instead.

The $K \rightarrow \pi$ transitions are given by:
\begin{equation}
{\cal A}(K^+ \rightarrow \pi^+)_{E,H} \, = \,
\frac{2 i}{f^2} \, (m_s +m_d) \, k^2 \, G_{E,H}^{(4)} \; ,
\end{equation}
\begin{equation}
{\cal A}(K^+ \rightarrow \pi^+)_K \, = \,
\frac{4 i}{f^2} \, m_u \, k^2 \, G_K^{(4)} \, ,
\end{equation}
which show that
$G_E^{(4)}(Q_6)$ and $G_H^{(4)}(Q_6)$ cannot be distinguished at this level
(to consider other off-shell meson-to-meson transitions does not solve
the problem). Thus,
from the calculation of
$K \rightarrow \pi$ transitions (see Fig. 3 for the relevant diagrams)
we can
at most determine $G_K^{(4)}(Q_6)$ and the sum $[G_E^{(4)}(Q_6) +
G_H^{(4)}(Q_6) ]$.
 We therefore have also to consider the $K \rightarrow 2 \pi$ amplitudes
at the quark level, and match them with the chiral lagrangian results:
\bea
{\cal A}(K^0 \rightarrow \pi^+ \pi^-)_E & = &
\frac{\sqrt{2}}{f^3} \, (m_s - m_d) \, m_{\pi}^2 \, G_E^{(4)} \, ,
\label{AE}\\
{\cal A}(K^0 \rightarrow \pi^+ \pi^-)_H & = &
\frac{\sqrt{2}}{f^3} \, \left[ (m_s + m_d) \, (m_K^2 - m_{\pi}^2)
   \right. \nnu \\
& & \! \! \! \! + \left. (m_u + m_d) \,(m_K^2 - m_{\pi}^2)
  +  (m_u - m_d) \, m_{\pi}^2 \right] G_H^{(4)} \, ,
\label{AH}\\
{\cal A}(K^0 \rightarrow \pi^+ \pi^-)_K & = &
\frac{2 \sqrt{2}}{f^3} \, \left[ (m_d - m_u) \, m_{\pi}^2  +
 m_d \,(m_K^2 - m_{\pi}^2)\, \right] G_K^{(4)} \, .
\label{AK}
\eea
 From \eq{AH} it appears that in order to
determine $G_H^{(4)}(Q_6)$ by calculating quark loops for $K \rightarrow 2
\pi$, we may keep the $m_s m_K^2$ terms only.
Moreover,
if the coefficients $G^{(4)}_{E,H,K}(Q_6)$ are of the same order of
magnitude, the term ${\cal{L}}^{(4)}_H$ will be the most important one.

To determine the coefficients  $G_{E,H,K}^{(4)}(Q_6)$, we calculate the
$K \rightarrow \pi$
and $K \rightarrow 2 \pi$ amplitudes due to $Q_6$ within the $\chi$QM.
For the $K \rightarrow \pi$ transitions,
in the leading factorizable limit, we obtain:
\bea
\langle \pi^+|Q_6|K^+ \rangle \, &=& 2 \, \langle
\pi^+|\overline{u}\gamma_5 d|0
\rangle  \langle 0|\overline{s} \gamma_5 u |K^+ \rangle  \nonumber \\
 & & -
2 \left[ \langle 0|\overline{d} d|0 \rangle + \langle 0|\overline{s} s|0
\rangle \right]
\langle \pi^+|\overline{s} d |K^+ \rangle
\eea
and for $K \rightarrow 2 \pi$:
\bea
\langle \pi^+ \pi^-|Q_6| K^0 \rangle \, &=&
2 \, \langle \pi^+|\overline{u}\gamma_5 d|0 \rangle
\langle \pi^-|\overline{s} u |K^0 \rangle
- 2  \langle \pi^+ \pi^-|\overline{d} d|0 \rangle  \langle 0|\overline{s}
\gamma_5 d |K^0 \rangle
\nonumber \\
& &
+ 2 \left[ \langle 0|\overline{s} s|0 \rangle \, - \, \langle 0|\overline{d}
d|0 \rangle
\right] \,  \langle \pi^+ \pi^-|\overline{s}\gamma_5 d |K^0 \rangle \, .
\eea

These equations contain some building blocks  which we
 calculate  within the $\chi$QM. The quark condensate
(Fig.~2) with mass insertions included is given by
\beq
\langle \overline{q} q \rangle \, = \,  \langle \overline{q} q \rangle ^{(0)}
  \, +  \, \frac{1}{2 M} ( m_s +  m_u)
\left[ \langle \overline{q} q \rangle ^{(0)}
\ +\ M f f_{\pi}^{(0)} \right]
\ +\ \langle \overline{q} q\rangle_G
%\ -\ \frac{1}{12M} \, \langle \frac{\alpha_s}{\pi}G^2 \rangle
\, .
\label{qqmass}
\eeq
where the last term represents the two-gluon condensate contribution (see
\eq{GGqq}).
As for the diagrams in Fig. 4, in the naive dimensional
regularization (NDR) scheme (anti-commuting $\gamma_5$ in $d\neq 4$),
we obtain:
\bea
 \vev{ 0|\,\overline{s} \gamma_5 u\,
|K^+(k)}_{\mbox{\scriptsize NDR}} &  =&  i \sqrt{2} \;
\left[ \frac{\vev{\overline{q} q}^{(0)}}{f} - k^2 \,
\frac{f_{\pi}^{(0)}}{2 M}
 \right. \nnu \\
& & \left.  +  (m_s + m_u) \left( f_{\pi}^{(0)}  +  3\, f
\frac{k^2}{ \Lambda_\chi^2} \right)
\right]
\, , \label{KvacuumNDR} \\
\langle \pi^+(p_+)|\,\overline{s} d\, |K^+(k)
\rangle_{\mbox{\scriptsize NDR}} \, &=&
\, - \frac{\langle \overline{q} q
\rangle ^{(0)}}{f^2}
 +  \frac{3 M}{2 \Lambda_\chi^2} P^2 \, +
\, \frac{q^2}{2 M} \left( f_+ \, \, - 3\, \frac{M^2}{\Lambda_\chi^2} \right)
 \nonumber \\
 & & - \, m_s \left[ \left( f_+  - 6\, \frac{M^2}{\Lambda_\chi^2} \right)
 - \frac{q^2
+q \cdot P}{2 \Lambda_\chi^2} \right]  \nnu \\
 & & -  2\, m_u  \left( f_+  +  3\, \frac{k^2}{\Lambda_\chi^2} \right)
\, ,     \label{KpiNDR}
\eea
where  $q=k-p_+$ and $P=k+p_+$, while $f_+ \equiv f^{(0)}_{\pi}/f = 1$ can
 be identified with the vector form factor at
zero momentum transfer $q$.
Similarly, in the 't Hooft--Veltman (HV) scheme (commuting $\gamma_5$ in
$d \neq 4$), we find:
\bea
 \vev{ 0|\,\overline{s} \gamma_5 u\, |K^+(k)}_{\mbox{scriptsize HV}} &  =&
  \vev{ 0|\,\overline{s} \gamma_5 u\, |K^+(k)}_{\mbox{\scriptsize NDR}}
\nnu \\
& &  +  i \sqrt{2}\ f\
\left[
12 \, \frac{M^3}{\Lambda_\chi^2}\left(1 - \frac{k^2}{6 M^2}\right)
\right. \nnu \\
& & \left. + 12  \, (m_s + m_u )  \frac{M^2}{\Lambda_\chi^2} \right]
\, , \label{KvacuumHV} \\
 \vev{ \pi^+ (p_+)|\,\overline{s} d\, |K^+(k)}_{\mbox{\scriptsize HV}} &  =&
  \vev{ \pi^+ (p_+)|\,\overline{s} d\, |K^+(k)}_{\mbox{\scriptsize NDR}}
 -  24\,
\frac{M^3}{\Lambda_\chi^2}
\, . \label{KpiHV}
\eea
In \eqs{KvacuumNDR}{KpiHV} we have introduced the chiral symmetry-breaking
scale
\beq
\Lambda_\chi \equiv 2\pi\ \sqrt{\frac{6}{N_c}}\ f = 0.83 - 1.0\ \mbox{GeV}\, ,
\label{lambda-chi}
\eeq
where the given range corresponds to assigning to $f$ the numerical values
of $f_\pi$ and $f_K$, as indication of $SU(3)$-breaking effects. Notice that
identifying $f^{(0)}_\pi = f$ in \eq{fqq}
leads to $f\sim \sqrt{N_c}$, which makes
$\Lambda_\chi$  independent of $N_c$.
As a consequence, the $N_c$ dependence of \eqs{KvacuumNDR}{KpiHV} resides
entirely in $f$ and $\vev{\overline{q}q}\sim N_c$.

The terms containing current quark masses are obtained from
mass-insertions due to ${\cal{L}}_{mass}$ in \eq{25} in the various quark
loop diagrams in Fig. 4. We
have discarded terms proportional to  $m_d$ because they are not
needed in determining $G_{E,H,K}^{(4)}(Q_6)$.
The  matrix element
$\langle \pi^+(p_+)|\overline{u}\gamma_5 d|0 \rangle$ is obtained from
 $\langle 0|\overline{s}\gamma_5 u |K^+(k) \rangle$
by obvious substitutions, while
$\langle \pi^+(p_+) \pi^-(p_-)|\overline{d} d|0 \rangle$
(where $p_+ + p_- = k$) can be derived from
\eq{KpiNDR} and (\ref{KpiHV}) by crossing ($k \rightarrow -p_-$).
Finally,
$\langle \pi^+ \pi^-|\overline{s}\gamma_5 d | K^0 \rangle$
is zero to this order~\cite{cheng,Don}.
Notice that these expressions
 are more general than those given by other authors~\cite{cheng,Don},
because our results are not  based on the divergence of the on-shell
current.

By using these matrix elements, we find that the constant term for the
$Q_6-$induced
$K \rightarrow \pi$ transition vanishes as it should according to
chiral symmetry~\cite{Don,DuPh,Orsay}. This cancellation is exact in NDR; in
the HV, it holds only
up to terms of order $1/\Lambda_\chi^4$, a warning about the relevance of
higher-order
loop effects. A typical example of such a high order
contribution would be any diagram in
Fig. 3 with an extra meson line connecting the two quark loops. For
consistency, we will always drop terms of order $1/\Lambda_\chi^4$
or higher in all numerical estimates.

In the NDR scheme, neglecting current quark masses which would amount to
a correction $\leq 10\%$,
the LO coefficient for the $Q_6$ matrix element is therefore
\begin{eqnarray}
G_{\underline{8}}^{(2)}(Q_6)  =  C_6 \, \frac{2 f^2}{M}
\langle \overline{q}q \rangle ^{(0)}\
 \left( f_+ - 6\, \frac{M^2}{\Lambda_\chi^2} \right)
\, ,
\label{G6NDR}
\end{eqnarray}
while the result in the HV scheme is
\beq
G_{\underline{8}}^{(2)}(Q_6) = C_6 \ \frac{2f^2}{M}
\vev{\overline{q}q}^{(0)}\
\left[ f_+  -  2\, \frac{M^2}{\Lambda_\chi^2} +
12 \frac{M^3f ^2}{\vev{\overline{q} q}^{(0)} \Lambda_\chi^2} \left(
1 + 4 \, \frac{M^2}{\Lambda_\chi^2}  \right)
 \right] \, . \label{G6HV}
\eeq
Then the relation
\beq
G_{\underline{8}}^{(2)}(Q_6) \equiv -16 \, C_6 \,  \frac{L_5}{f^2}
\left[\langle \overline{q}q \rangle ^{(0)} \right]^2
\, ,
\label{L5def}
\eeq
defines the NLO chiral parameter $L_5$
in agreement with refs.~\cite{ERT,BBdR}.
In passing,
let us remark that the leading non-zero matrix element of $Q_6$ is
 next-to-leading in the framework of the chiral perturbation
 theory since it is proportional to the coefficient $L_5$ of
$\Tr [(\Sigma^{\dagger} \, {\cal{M}}_{q} \, +
 \; {\cal{M}}_{q}^{\dagger} \,\Sigma )  \,
 D^{\mu} \Sigma^{\dagger} D_{\mu} \Sigma ]$
in ${\cal{L}}^{(4)}_s$.

In the $\chi$QM, to order $M^2/\Lambda_\chi^2$, we find
\beq
L_5^{\mbox{\scriptsize NDR}} = -\frac{f^4}{8 M \vev{\overline{q}q}}
\left( f_+ - 6 \,\frac{M^2}{\Lambda_\chi^2} \right)
\, , \label{l5NDR}
\eeq
as in ref.~\cite{BBdR}, and
\beq
L_5^{\mbox{\scriptsize HV}} = -\frac{f^4}{8 M \vev{\overline{q}q}}
\left( f_+ -  2\, \frac{M^2}{\Lambda_\chi^2} +
12 \frac{M^3f ^2}{\vev{\overline{q} q}^{(0)} \Lambda_\chi^2} \right)
\, , \label{l5HV}
\eeq
both sensitive functions of $M$.
Let us remark that for
\beq
L_5 =\frac{1}{4}\,
\frac{f^2_\pi}{\Lambda_\chi^2}\left(\frac{f_\pi}{f_K}\right)^4
\label{l5Bu}
\eeq
\eq{L5def} used in \eq{A2(Q6)} gives the usual $1/N_c$ matrix
element~\cite{1/N}. In that scheme we therefore have
\bea
L_5 & = & 1.4\times 10^{-3}
\qquad \qquad \qquad \qquad \mbox{for} \quad \Lambda_\chi = 0.83 \:
\mbox{GeV} \nnu
\\
L_5 & = & 1.0\times 10^{-3}
\qquad \qquad \qquad \qquad \mbox{for} \quad \Lambda_\chi = 1.0 \:
\mbox{GeV}
\, . \label{range}
\eea
A scale-dependent value for $L_5$ can be extracted in $\chi$PT
from the ratio $f_K/f_\pi$ where one finds~\cite{l5exp}
\beq
\mbox{$L_5(m_\rho) = (1.4 \pm 0.5) \times 10^{-3}$} \ ,
\label{l5-rho}
\eeq
%For the corresponding value at $\mu=1$ GeV we obtain
%\beq
%\mbox{$L_5$(1 GeV)$ = (0.9 \pm 0.5) \times 10^{-3}$} \ ,
%\label{l5-1}
%\eeq
%which should be used for comparison with \eqs{l5NDR}{l5Bu}.
which we use for comparison with \eqs{l5NDR}{l5Bu}.

In Table 2 we have collected the results of the $\chi$QM predictions
for $L_5$ in the two schemes as we vary $M$.
A constraint of approximately 3$\sigma$ from the central value in \eq{l5-rho}
gives
\beq
140\ \mbox{MeV}\ltap M \ltap 340\ \mbox{MeV}
\label{range2} \, ,
\eeq
where we have taken for the scale dependent quark condensate the expression
in \eq{qq(mu)}, and assumed a perturbative running for
$\mu \geq 1$ GeV.
Recall however that for $M > 300$ MeV $M^4/\Lambda_\chi^4$ corrections become
relevant and may affect the smaller values of $L_5$ listed in Table 2.

Moving on to the NLO contributions, by using \eqs{AE}{AK} we find
in the NDR scheme:
\bea
G_E^{(4)}(Q_6)  &=&  - \, 2 \, C_6 \, \langle \overline{q}q \rangle ^{(0)}
\left(\frac{f^2}{\Lambda_\chi^2}
- \frac{3 f^4}{2 M \vev{\overline{q}q}^{(0)}}  \right) \, , \\
G_K^{(4)}(Q_6)  &  =  & 3\ C_6 \, \langle \overline{q}q \rangle ^{(0)}
\left(\frac{f^2}{\Lambda_\chi^2}
+ \frac{f^4 f_+^2}{6 M \vev{\overline{q}q}^{(0)}}  \right) \, ,
\label{GE-NDR}
\eea
whereas in the HV scheme:
\bea
G_E^{(4)}(Q_6)  &=&  - \, 2 \, C_6 \, \langle \overline{q}q \rangle ^{(0)}
\left[\frac{f^2}{\Lambda_\chi^2}
- 6 \frac{M^3 f^4}{\vev{\bar{q}q}^{(0)}\Lambda_\chi^4}
- \frac{5 f^4}{ M \vev{\bar{q}q}^{(0)}}
\left( 1 + 12 \frac{M^2}{\Lambda_\chi^2} \right)  \right] \, , \\
G_K^{(4)}(Q_6)  &  =  & 3\ C_6 \, \langle \overline{q}q \rangle ^{(0)}
\left[\frac{f^2}{\Lambda_\chi^2} \left( 1
+ 12 \frac{M^3 f^4}{\vev{\bar{q}q}^{(0)}\Lambda_\chi^4}\right) \nnu
\right. \\
& &  \hspace{2.5cm} \left. - \frac{f^4}{6 M \vev{\bar{q}q}^{(0)}}
\left( f_+  + 4 \frac{M^2}{\Lambda_\chi^2} \right)
\left( f_+  + 12 \frac{M^2}{\Lambda_\chi^2} \right)  \right]
\, .
\eea

The most important contribution to $K \rightarrow 2 \pi$ comes from
the third coefficient $G_H^{(4)}(Q_6)$, since the corresponding amplitude
is enhanced by a factor $m_K^2/m_\pi^2$ compared to the other two
(see \eqs{AE}{AK}). In the NDR scheme we find:
\beq
G_H^{(4)}(Q_6) \ \ =\ \ - 7 \, C_6 \, \langle \overline{q}q \rangle ^{(0)}
\left[ \frac{f^2}{\Lambda_\chi^2}
- \frac{f^4}{7 M \vev{\bar{q}q}^{(0)}}
\left( f_+ - 6 \frac{M^2}{\Lambda_\chi^2} \right)  \right] \, ,
\eeq
whereas the HV result is:
\bea
G_H^{(4)}(Q_6) &  = &    - 7 \, C_6 \, \langle \overline{q}q \rangle ^{(0)}
\left[ \frac{f^2}{\Lambda_\chi^2}
+ \frac{156}{7} \frac{M^3 f^4}{\vev{\bar{q}q}^{(0)}\Lambda_\chi^4} \nnu
\right. \\
& &  \hspace{2.5cm} \left. - \frac{f^4}{7 M \vev{\bar{q}q}^{(0)}}
\left( f_+  - 6 \frac{M^2}{\Lambda_\chi^2} \right)
\left( 1 + 12 \frac{M^2}{\Lambda_\chi^2} \right)  \right]
\, . \label{GH-HV}
\eea

Recently an NLO analysis of the hadronic matrix elements $Q_{1-10}$ has
appeared~\cite{fajf}, which makes use of effective scalar-meson exchange.
The result of ref.~\cite{fajf} for the operator $Q_6$
does not contain terms corresponding
to ${\cal{L}}^{(6)}_s$ (with two derivatives and current masses squared),
which contribute to \eqs{AE}{AK}.

We see
that $G_{E,H,K}^{(4)}(Q_6)$ are formally suppressed by terms
proportional to either
$ f^2/\Lambda_\chi^2$ or $ f^4/(M \vev{\bar{q}q}^{(0)})$
with respect to $G^{(4)} (Q_{11})$.
However, because of the numerical factors in front,
the NLO $G^{(4)}(Q_6)$ coefficients turn out to be
numerically of the same size as $G^{(4)} (Q_{11})$.
Since  the
amplitude due to ${\cal{L}}^{(4)}(Q_{11})$
is effectively suppressed by a factor
$m_\pi^2/m_K^2$  with respect
to the amplitudes obtained from  ${\cal{L}}^{(2)}(Q_{6})$ and
${\cal{L}}_H$,
the contribution of $Q_{11}$ to $\varepsilon '/\varepsilon$
turns out to be generally
more than one order of magnitude smaller than
these NLO corrections (see Table 3). The same
can be said of other subleading operators such as $Q_3$, $Q_5$ and $Q_7$
(see \eq{Q1-10} below for the definitions, and Tables 7--9).

In Table 3 we have reported the weights
of the NLO matrix elements of $Q_6$
and $Q_{11}$ relative to the LO $Q_6$ amplitude, as computed
in the $\chi$QM.
These estimates are robust as they do not sensibly depend on the detailed
values of $\Lambda_{QCD}$
and $\Lambda_\chi$.

\section{Scheme dependence of $\langle Q_6 \rangle_{\underline{8}}$ and
$\langle Q_8 \rangle_{\underline{27}}$}

The Wilson coefficients computed by means of the  renormalization group
equations depend, at the NLO, on whether the NDR or
the HV prescriptions are used in treating $\gamma_5$ in $d$
dimension. At the same time, the matrix elements of the relevant
operators do not have any scheme dependence, at least if computed
by $1/N_c$ techniques. As a consequence, there remains
a $\gamma_5$-scheme dependence
which increases the uncertainties in the final result. The discrepancy between
the HV and NDR result goes from $30 \%\: (\Lambda_{QCD} = 200$ MeV) to $40 \%
\: (\Lambda_{QCD} = 400$ MeV) at $m_t = 130$ MeV,
becoming much worse in the ``superweak'' regime at $m_t = 170$ GeV,
where it goes
from $70 \% \: (\Lambda_{QCD} = 200$~MeV) to $80 \% \:
(\Lambda_{QCD} = 400$ MeV)~\cite{Monaco}.

A potentially nice feature of  the $\chi$QM is that it
makes it  possible to compute
also the matrix elements in both schemes.
In \eqs{G6NDR}{G6HV} we have computed the matrix element for the operators
$Q_6$ to the leading order in the chiral expansion.
The next most relevant contribution comes from the electroweak penguin
operator $Q_8$ (\eq{Q8}). This operator can be written as
\beq
Q_8 = \frac{3}{2} e_d \cdot Q_6\ \ +\ \
\frac{3}{2}(e_u-e_d)\ \overline{s}_{\alpha}
\gamma_\mu ( 1 - \gamma_5 ) d_{\beta}
                 \ \  \overline{u}_{\beta}
\gamma^\mu ( 1 + \gamma_5 ) u_{\alpha}\ ,
\label{Q8decomp}
\eeq
where $e_u = -2 e_d = 2/3$, so as to obtain
\beq
\vev{ \pi^+ \pi^- | Q_8 | K^0}_{\underline{8}} \ \ =\ \
\vev{ \pi^0 \pi^0 | Q_8 | K^0}_{\underline{8}} \ \ =\ \
-\frac{1}{2}\
\vev{ \pi^+ \pi^- | Q_6 | K^0}_{\underline{8}} \ ,
\label{Q8-8}
\eeq
and, after a Fierz rearrangement and factorization,
\bea
\vev{ \pi^+ \pi^- | Q_8 | K^0}_{\underline{27}} & = &
 3\
\vev{\pi^+ | \bar{u} \gamma_5 d | 0} \vev{\pi^- | \bar{s} u | K^0} \ ,
\label{Q8-27p} \\
\vev{ \pi^0 \pi^0 | Q_8 | K^0}_{\underline{27}} & = & 0\ .
\label{Q8-27o}
\eea
The leading order amplitude corresponding to \eq{Q8-27p} is therefore given
by
\beq
{\cal A}( K^0 \rightarrow \pi^+ \pi^- ;\, Q_8)_{\underline{27}}
\ \ =\ \  C_8 \frac{3\sqrt{2}}{f^3}
\Bigl[ \vev{\overline{q}q} ^{(0)} \Bigr] ^2
\label{G8NDR}
\eeq
in the NDR scheme, and
\bea
{\cal A}( K^0 \rightarrow \pi^+ \pi^- ;\, Q_8)_{\underline{27}}
& = &  C_8 \frac{3\sqrt{2}}{f^3} \Bigl[
 \vev{\overline{q}q} ^{(0)} \Bigr] ^2 \times  \nnu \\
& &  \hspace{1cm} \left( 1
+ 12 \frac{M^3 f^2}{\vev{\overline{q}q} ^{(0)}\Lambda_\chi^2} \right)
\left( 1
+ 24 \frac{M^3 f^2}{\vev{\overline{q}q} ^{(0)}\Lambda_\chi^2} \right)
\label{G8HV}
\eea
in the HV scheme.
In \eqs{G8NDR}{G8HV} we have kept only the leading momentum-independent terms.
The momentum-dependent corrections are at the 10\% level.

 In a toy model for $\varepsilon '/\varepsilon$ in
which we use the full mixing to determine the Wilson coefficients, but
 keep only the contribution of the
two leading operators $Q_6$ (for the $I=0$ amplitude) and $Q_8$ (for
$I=2$), we can compare the scheme dependence in the $1/N_c$
approach---where it
only appears in the Wilson coefficients---and in the $\chi$QM, for
different choices of $M$.

We have taken the Wilson coefficients
at $\mu=$ 1 GeV ($\mu=$ 1.4 GeV) for $m_t=170$~GeV and multiplied them
by the corresponding  matrix elements,
according to \eq{eps} in the next section. We have used
\bea
\langle 2 \pi , \: I=0 |\, Q_{i} \, | K^0 \rangle & = &
\sqrt{\frac{2}{3}} \: \langle \pi^+ \pi^- | Q_{i} | K^0 \rangle
\ \ +\ \
\sqrt{\frac{1}{6}} \: \langle \pi^0 \pi^0 | Q_{i} | K^0 \rangle \ ,
\label{Qi-I0} \\
\langle 2 \pi , \: I=2 |\, Q_{i} \, | K^0 \rangle & = &
\sqrt{\frac{1}{3}} \: \Bigl[ \langle \pi^+ \pi^- | Q_{i} | K^0 \rangle
\ \ -\ \
\langle \pi^0 \pi^0 | Q_{i} | K^0 \rangle \Bigr] \ .
\label{Qi-I2}
\eea
In our case
\bea
\langle 2 \pi , \: I=0 |\, Q_{6} \, | K^0 \rangle  & = &
\sqrt{\frac{3}{2}} \: \langle \pi^+ \pi^- | Q_{6} | K^0
\rangle_{\underline{8}}  \, , \\
\langle 2 \pi , \: I=2 |\,  Q_{8}\, | K^0 \rangle  & = &
\sqrt{\frac{1}{3}} \: \langle \pi^+ \pi^- | Q_{8} | K^0
\rangle_{\underline{27}}  \, .
\label{Q8-I2}
\eea
Consistently to our approximation, we neglect the $I=0$ contribution of $Q_8$:
\beq
\langle 2 \pi , \: I=0 |\, Q_{8} \, | K^0 \rangle
\ \ =\
-\ \frac{1}{2}\ \langle 2 \pi , \: I=0 |\, Q_{6} \, | K^0 \rangle
%\sqrt{\frac{3}{2}} \: \langle \pi^+ \pi^- | Q_{8} | K^0
%\rangle_{\underline{8}}
\ +\
\sqrt{\frac{2}{3}} \: \langle \pi^+ \pi^- | Q_{8} | K^0
\rangle_{\underline{27}} \ ,
\label{Q8-I0}
\eeq
since it is suppressed in $\varepsilon'/\varepsilon$
by a factor $\sim 1/20$ ($\Delta I = 1/2$ rule)
with respect to the $I=2$ contribution of \eq{Q8-I2}.
Notice that, at the zeroth order in momentum expansion,
$\vev{\pi^+ \pi^- | Q_{6} | K^0} = 0$
and, as a consequence, the last term in \eq{Q8-I0} is leading.

The matrix elements in the $\chi$QM are computed using
the scale-dependent quark condensate
\beq
\vev{\overline{q}q} (\mu ) = - \frac{f_K^2 m_K^2}{m_s(\mu)+m_d(\mu)}
 \, .
\label{qq(mu)}
\eeq
Following ref. \cite{Monaco}, we take $m_s(1.4\ \mbox{GeV})= 150\ \mbox{MeV}$
and $m_d(1.4\ \mbox{GeV})= 8.0\ \mbox{MeV}$. The values at other scales are
obtained using the NLO evolution of QCD. In particular, for
$\Lambda_{QCD} = 300\ \mbox{MeV}$
one has $m_s(1.0\ \mbox{GeV})= 176\ \mbox{MeV}$ and
$m_d(1.0\ \mbox{GeV})= 9.4\ \mbox{MeV}$.

Tables 4 and 5 summarize our findings
for different  values of $\Lambda^{(4)}_{QCD}$.
In using \eq{G6HV} and \eq{G8HV} we
have dropped all terms of order $1/\Lambda_\chi^4$ for consistency and limited
ourselves to values of $M < 220$ MeV. As explained in the
beginning, we cannot trust our results
beyond order $1/\Lambda_\chi^2$ because we have neglected
higher-loop corrections
with meson exchange that give $1/\Lambda_\chi^4$ contributions.

In the range considered,
the $\gamma_5$-scheme dependence is indeed dramatically
reduced. For instance, taking
$\Lambda^{(4)}_{QCD} = 300$ MeV,
$\Delta$---defined as the difference between the HV and the NDR
results divided by the HV one---is below $10 \%$ for $M
\simeq 120-150$ MeV ($\mu = 1.4$ GeV) and for $ M \simeq 120-170$ MeV
($\mu = 1$ GeV),
when $\Lambda_\chi = 830$ MeV; similarly, $\Delta$ is below $10 \%$
for $M
\simeq 120-180$ MeV ($\mu = 1.4$~GeV) and for $ M \simeq 150-190$ MeV
($\mu = 1$ GeV),
when $\Lambda_\chi = 1$~GeV.

Another, and more restrictive,
reading of the same tables puts together all data for
different $\Lambda_{QCD}$'s. In this case, stability is achieved for $M
\simeq 120-150$ MeV ($\mu = 1.4$~GeV) and for $ M \simeq 150$ MeV
($\mu = 1$ GeV), when $\Lambda_\chi = 830$ MeV; similarly, $\Delta$ is below
 $10 \%$
for $M
\simeq 120-170$ MeV ($\mu = 1.4$ GeV) and for $ M \simeq 190$ MeV
($\mu = 1$ GeV),
when $\Lambda_\chi = 1$ GeV.

To reach values of $M$ larger than its central value
and closer to it,
we would have to include higher-loop  corrections.
The preliminary nature of our analysis needs
hardly be stressed as only two out of eleven operators have been
considered.
Even though $Q_6$ and $Q_8$ induce the most relevant contributions,
we expect that a complete estimate of $\varepsilon'/\varepsilon$
would result in a more stable range of values of $M$
for which the $\gamma_5$-scheme and $\mu$
dependences are reduced. This will also give us confidence on the size
of $\varepsilon'/\varepsilon$ that is obtained.

\section{``NLO'' study of $\varepsilon'/\varepsilon$ with dipole operators}

In this section we discuss the impact of the dipole gluon penguin
$Q_{11}$ on present estimates of $\varepsilon'/\varepsilon$. The contribution
of $Q_{11}$ is given by setting $\vev{\bar{q}q}_G$ at the rather conservative
value of $-(275 \, \mbox{MeV})^3$ (see Table 1).

Since a satisfactory calculation of $\varepsilon'/\varepsilon$ in the
$\chi$QM is missing beyond the leading factorizable order
(the study is under way),
we resort to the
$1/N_c$ analysis of ref.~\cite{Monaco} for the ten traditional operators.
We present our results in tables that show
a detailed anatomy of the contributions of the different operators,
and can be used as reference for future developments.

Because of the many ingredients involved in the
calculation of $\varepsilon'/\varepsilon$, it is useful
to briefly recall the theoretical inputs used.
The effective lagrangian for $|\Delta S| = 1$ transitions can be written,
for $\mu<m_c$,
as~\cite{Monaco} \beq
{\cal L}_{\Delta S = 1} = -
\frac{G_F}{\sqrt{2}} \lambda_u  \sum_i \Bigl[
z_i(\mu) + \tau y_i(\mu) \Bigr] Q_i (\mu) = \sum_{i} C_i(\mu) Q_i(\mu)
 \, . \label{ham}
\eeq
In the previous equation, $\lambda_i \equiv
V_{id}V^*_{is}$, where $V$ is the Kobayashi-Maskawa (KM) matrix, and
$\tau \equiv  - \lambda_t/\lambda_u$.
The Wilson coefficients $z_{1,2}(\mu)$
run from $m_W$ to $m_c$ via the corresponding $2\times 2$ sub-block
of the $10\times 10$ anomalous-dimension matrices, while $z_i(\mu)=0$ for
$i=3-10$. From $\mu=m_c$ down, as the charm-induced penguins come into play,
all $z_i(\mu)$ evolve, given the proper matching conditions,
with the full anomalous-dimension matrices.
The Wilson coefficients $v_i(\mu)$ ($y_i (\mu) = v_i (\mu) - z_i (\mu)$)
arise at $m_W$ due to integration of the $W$ and top quark fields.
They coincide with $z_i(\mu)$ for $i=1,2$,
the information about the
top quark being encoded in the $i=3-10$ components.

The list of the effective operators $Q_{i}$ ($i=1-10$) is reported
in refs.~\cite{Monaco,Roma}, whose notation
we  follow closely and where the reader may find a complete
discussion of the basic tools.
For convenience we report here the ten operators
usually considered
\beq
\begin{array}{rcl}
Q_{1} & = & \left( \overline{s}_{\alpha} u_{\beta}  \right)_{\rm V-A}
            \left( \overline{u}_{\beta}  d_{\alpha} \right)_{\rm V-A}
\, , \\[1ex]
Q_{2} & = & \left( \overline{s} u \right)_{\rm V-A}
            \left( \overline{u} d \right)_{\rm V-A}
\, , \\[1ex]
Q_{3,5} & = & \left( \overline{s} d \right)_{\rm V-A}
   \sum_{q} \left( \overline{q} q \right)_{\rm V\mp A}
\, , \\[1ex]
Q_{4,6} & = & \left( \overline{s}_{\alpha} d_{\beta}  \right)_{\rm V-A}
   \sum_{q} ( \overline{q}_{\beta}  q_{\alpha} )_{\rm V\mp A}
\, , \\[1ex]
Q_{7,9} & = & \frac{3}{2} \left( \overline{s} d \right)_{\rm V-A}
         \sum_{q} e_q \left( \overline{q} q \right)_{\rm V\pm A}
\, , \\[1ex]
Q_{8,10} & = & \frac{3}{2} \left( \overline{s}_{\alpha}
                                                 d_{\beta} \right)_{\rm V-A}
     \sum_{q} e_q ( \overline{q}_{\beta}  q_{\alpha})_{\rm V\pm A}
\, ,
\end{array}  \label{Q1-10}
\eeq
where $\alpha$, $\beta$ denote color indices ($\alpha,\beta
=1,\ldots,N_c$) and $e_q$  are quark charges. Color
indices for the color singlet operators are omitted.
$(V\pm A)$ refer to
$\gamma_{\mu} (1 \pm \gamma_5)$.
We recall that
$Q_{1,2}$ stand for the $W$-induced current--current
operators, $Q_{3-6}$ for the
QCD penguin operators and $Q_{7-10}$ for the electroweak penguin (and box)
ones.

Not all the operators in \eq{Q1-10} are independent.
For $\mu < m_c$, having integrated out the charm quark,
we have
\begin{eqnarray}
Q_4    &=& Q_3 + Q_2 - Q_1 \, ,\nnu \\
Q_9    &=& \left(3 Q_1 - Q_3\right)/2 \, ,
\label{Q4910} \\
Q_{10} &=& Q_2 + \left(Q_1 - Q_3\right)/2 \; .  \nnu
\end{eqnarray}
Note that these relations hold in the HV scheme,
but they may receive additional contributions in other schemes
since Fierz transformations have been used in obtaining them.

Together with this basis, which closes under QCD and QED renormalization,
one should a-priori consider two additional dimension-five operators:
$Q_{11}$ (\eq{Q11}) and
\beq
Q_{12} \simeq  \frac{ee_d}{16\pi^2}\, m_s \: \overline{s}\, \sigma_{\mu\nu}
 F^{\mu\nu} (1-\gamma_5 )\, d  \ ,
\label{Q12}
\eeq
which account for the magnetic and electric dipole part of,
respectively, the QCD and electromagnetic penguin operators.
In  \eq{Q12} $e_d = -1/3$
is the charge of the down quarks.

Actually, in \cite{BFG} we argued that the hadronic matrix element of
the electromagnetic operator (\ref{Q12})
is negligible and, accordingly, even if we keep the operator in our
basis for the Wilson coefficients, we will put its contribution to zero in the
end.

Since Im$(\lambda_u) =0$ according to the standard conventions,
 the short-distance component of $\varepsilon'/\varepsilon$
is determined by the Wilson coefficients $y_i$.
Following the approach of
ref.~\cite{Monaco}, $y_1(\mu)=$ $y_2(\mu)=0$ and the effect of $Q_{1,2}$
appears
only through the linearly-dependent operators $Q_{4,9,10}$.

The lagrangian in \eq{ham} yields \cite{Monaco}
\beq
\frac{\varepsilon '}{\varepsilon} = 10^{-4} \left[ \frac{ \mbox{Im}\, \lambda
_t}{1.7 \times 10^{-4}} \right] \Bigl[ P^{(1/2)} - P^{(3/2)} \Bigr] \, ,
\label{eps}
 \eeq
where
\bea
 P^{(1/2)} & = & r\ \sum y_i \, \langle 2 \pi,\ I=0 |\, Q_i \, |  K^0 \rangle
\left(1 - \Omega_{\eta +\eta'} \right) \\
 P^{(3/2)} & = & \frac{r}{\omega} \sum y_i \, \langle 2 \pi,\ I=2 |\,  Q_i
\, | K^0 \rangle  \ .
\eea
We take, as input values for
the relevant quantities, the central values given in appendix~C of
ref. \cite{Monaco}. This allows us to reproduce, in the ten-operator case,
the central values of the results
given in appendix B of ref.~\cite{Monaco}.
In particular, we take
\beq
r = 1.7\frac{\omega G_F}{2\mod{\epsilon}\Re{A_0}} \simeq 594 \ \mbox{GeV}^{-3},
\qquad \omega = 1/22.2\ , \qquad \Omega_{\eta + \eta'} = 0.25 \ ;
\eeq
$\Im{\lambda_t}$ is determined from the experimental value of
$\varepsilon$ as an interpolating function of $m_t$.
Its central value, given
the KM phase $\delta_{KM}$ in the first or second quadrant, is given by
\beq
\Im{\lambda_t} \simeq 2.77\times 10^{-4}\ x_t^{-0.365}\ \ \ (\mbox{first
quadrant})\ ,
\label{Imlamti1}
\eeq
and
\beq
\Im{\lambda_t} \simeq 2.19\times 10^{-4}\ x_t^{-0.47}\ \ \ \mbox{(second
quadrant})\ ,
\label{Imlamti2}
\eeq
where $x_t = m_t^2/m_W^2$.

 The value of the Wilson coefficients $y_{11}$ and $y_{12}$ at
the hadronic scale of 1 GeV can be found by means of the renormalization
group equations. Denoting generically  the vector of
Wilson coefficients by $\vec C(\mu)$, its scale dependence is governed by

\beq
\left[\mu\frac{\partial}{\partial\mu}
+ \beta(g)\frac{\partial}{\partial g}\right]
\vec C \left( \frac{m_W^2}{\mu^2},g^2,\alpha \right) \ \ =\ \
\hat\gamma^T(g^2,\alpha)\ \vec C \left( \frac{m_W^2}{\mu^2},g^2,\alpha
\right) \, ,
\label{RGE}
\eeq
where $\beta(g)$ is the QCD beta function and $\alpha$ the electromagnetic
coupling (the running of $\alpha$ is being neglected).
At the NLO we have
\beq
\hat\gamma(g^2,\alpha) =
   \frac{\alpha_s}{4\pi}\left(\hat\gamma_s^{(0)}
 + \frac{\alpha_s}{4\pi}\hat\gamma_s^{(1)}\right)
 + \frac{\alpha}{4\pi}\left(\hat\gamma_e^{(0)}
 + \frac{\alpha_s}{4\pi}\hat\gamma_{se}^{(1)}\right)\ ,
\label{gammas}
\eeq
where $\hat\gamma_s^{(0)}$
and $\hat\gamma_e^{(0)}$
govern the leading QCD and the
electromagnetic running respectively.
The anomalous-dimension matrices labelled with $(1)$
refer to the NLO running ($O(\alpha^2)$ effects are
neglected).

In order to include all  available
NLO effects in the evaluation of
$\varepsilon'/\varepsilon$, we follow the analysis
described in ref.~\cite{Monaco}.
The NLO $10\times 10$ mixing matrices for the operators
$Q_{1-10}$ can be found in
refs.~\cite{Monaco,Roma}.
Concerning the dipole operators,
the leading-order matrix of the strong anomalous
dimensions of $Q_{11}$ and $Q_{12}$ and their QCD-induced mixing with
$Q_{1-6}$
can be borrowed from the existing calculations for the $b\to s\gamma$
decay \cite{qcdbsgamma} (recent discussions are given in
ref.~\cite{nlobsgamma}).
In fact, by replacing $s\to b$ and $d\to s$ in \eqs{Q1-10}{Q12}
we obtain the operator basis, which should be
considered for a complete NLO analysis of $b\to s\gamma$.

While the $10 \times 10$ part of the anomalous dimension matrices
(\ref{gammas}) is identical to that used in refs.~\cite{Monaco,Roma},
two extra columns and rows have to be added to represent the mixing of the
first ten operators with the two new ones, which takes place first at the
two-loop level.
We have taken for all two-loop anomalous dimensions
the HV scheme results~\cite{Monaco,Roma}.
In this way,
no finite additional contributions to the renormalization of
$y_{11}$ and $y_{12}$ arise at the various quark thresholds
(for a discussion see Misiak in ref.~\cite{qcdbsgamma}
and ref.~\cite{C11schemeind}).
The explicit expression of $\hat\gamma_s^T$ is reported for instance
in ref.~\cite{BFG}.
We just recall that whereas the evolution of the dipole Wilson
coefficients $C_{11}$ and $C_{12}$ is substantially affected
by the mixings with $Q_{1-10}$, the Wilson coefficients $C_{1-10}$
remain unaffected by the presence of the dipole operators
($Q_{1-10}$ close under QCD and QED renormalization).
The two-loop mixings of
$Q_{11}$ and $Q_{12}$ with the electroweak penguins ($Q_{7-10}$),
not yet given in the literature,
can be easily derived from those with the gluon penguins ($Q_{3-6}$).
We have verified that their effect on the running of $C_{11}$ and $C_{12}$
is negligible ($< 1\%$) and therefore they can be safely set to zero
(and, by extension, they can also be set equal to zero in $b \rightarrow s
\gamma$).

The complete NLO analysis would require computing, among other things, the
three-loop mixings between the dipole operators and the first ten
(quite a task!).
The lack of knowledge on these entries introduces an uncertainty
on the dipole Wilson coefficients which can be as large as 50\%
(see the analogous discussion for the $b\to s\gamma$ inclusive decay
in ref. \cite{nlobsgamma}).
However, for what concerns $\varepsilon'/\varepsilon$ this uncertainty is
diluted over many contributions, and it is certainly not as relevant as
our ignorance of the hadronic matrix elements.
The results presented here are obtained by adding
two rows and two columns of zeros
to the  $10\times 10$ electromagnetic and NLO
anomalous dimension matrices, which can be found in refs.~\cite{Monaco,Roma}.
As a consequence,
the contribution of $Q_{11}$ to $\varepsilon'/\varepsilon$ takes into account
only the ``LO'' (two-loop) QCD effects.

The expressions for the initial Wilson coefficients $v_{1-10}\,(m_W)$
can be found for instance
in ref.~\cite{Monaco}.
For what concerns the new coefficients $v_{11,12}\,(m_W)$ we have
\beq
v_{11}(m_W) =
-E'(m_t^2/m_W^2) \qquad \mbox{and} \qquad v_{12}(m_W) = -D'(m_t^2/m_W^2)/e_d
\, ,
\eeq
where
\bea
E'(x) & = & \frac{3x^2 }{2 (1 - x)^4} \ln x
-\frac{x^3 - 5 x^2 -2 x}{4 (1 - x)^3} \\
D'(x) & = & \frac{x^2 (2 - 3 x)}{2 (1 - x)^4} \ln x
-\frac{8 x^3 + 5 x^2 -7 x}{12 (1 - x)^3} \, .
\eea

In Table 6 we report the HV results we have
obtained for $z_{1-12}$ ($\mu=1$ GeV)
and $y_{3-12}$ ($\mu=1$ GeV) (recall that $y_{1,2}\,(\mu)=0$)
compared with their initial values, for $m_t=170$ GeV and for
$\Lambda_{QCD}^{(4)} = $ 200, 300 and 400 MeV.
We fully agree on the values of the renormalized coefficients
for the first ten operators with ref. \cite{Monaco}.

Let us remark that
the effect of operator mixing induces a
renormalization
on $v_{11,12}$ ($= y_{11,12}\, + \, z_{11,12}$)
which is a factor of 4--5 larger than that
induced by multiplicative running alone (which roughly reduces
by a factor of two the initial Wilson coefficients).

In order to discuss the effect of the various operators in
determining the size of $\varepsilon'/\varepsilon$,
we need a consistent estimate of
the relevant hadronic matrix elements.
For the operators $Q_{1-10}$, we follow the
strategy of ref.~\cite{Monaco}
where the various matrix elements are evaluated by means of
the $1/N_c$ expansion and soft-meson methods. Overall
coefficients $B_i^{(1/2)}$ and $B_i^{(3/2)}\ (i = 1-10)$
parametrize our level of ignorance of their normalization scale, scale
dependence and the approximation inherent in  the method.
The matrix elements of $Q_1$ and $Q_2$ can however be determined
phenomenologically from the experimental values of $\Re{A_0}$
and $\Re{A_2}$, so as to reproduce the $\Delta I= 1/2$ rule.
In particular, in ref.~\cite{Monaco} it is found that at $\mu=m_c$,
$B_{2,NLO}^{(1/2)}\approx 6.3$ in the HV
scheme, which is about three times larger
that the $1/N_c$ result. Related to this coefficient is the value
of $B_{1,NLO}^{(1/2)}$, which we find equal to 20.2, 13.5 and 8.1 for
$\Lambda_{QCD}^{(4)}=$ 200, 300 and 400 GeV, respectively. Correspondingly,
$B_{1,NLO}^{(3/2)}=0.45$, 0.46 and 0.48.
The large deviations from unity of these effective coefficients
gives us a gauge of
our lack of understanding of the $\Delta I= 1/2$ rule within the
$1/N_c$ expansion
(better results are achieved within the $\chi$QM \cite{PdeR91}, leaving
deviations at most of a factor 2--3).

Further relations among
other coefficients are advocated in ref.~\cite{Monaco},
depending on the relevance and the role of the various
operators, so as to reduce, in the ten-operator case, the description
of the hadronic sector to two effective parameters: $B_{6}^{(1/2)}$ and
$B_{8}^{(3/2)}$, whose leading $1/N_c$ value is 1.

The inclusion of $Q_{11}$ and $Q_{12}$ requires three additional effective
parameters: $B_{11}^{(1/2)}$, $B_{12}^{(1/2)}$ and $B_{12}^{(3/2)}$.
For $Q_{11}$ we
use our result (\ref{2.13}), where for $\vev{\overline{q} q}_G$ we take
a typical value from Table 1, namely $-(275$ MeV)$^3$.

Since the determination of
$B_{1}$ and $B_{2}$ is best achieved at $\mu=m_c$ \cite{Monaco},
all
the hadronic matrix elements are assumed to be evaluated at that scale
and renormalized down to 1 GeV via their anomalous-dimension matrix.
We proceed analogously, by setting $B_{11}^{(1/2)}=1$ and, as we neglect
$Q_{12}$,
$B_{12}^{(1/2)}=B_{12}^{(3/2)}=0$ at $\mu=m_c$ and using the
$12\times 12$ QCD and electromagnetic evolution matrices
to evolve all the hadronic matrix elements to the 1 GeV scale.

Since the anomalous-dimension matrices, which govern the
evolution of the hadronic matrix elements, are the transpose
of those evolving the Wilson coefficients, we now find  that
the presence of $Q_{11,12}$ affects, from $\mu=m_c$ down, the renormalization
of the first ten operators. On the other hand, the evolution
of $Q_{11,12}$ is determined solely by their $2\times 2$
anomalous-dimension matrix, which implies that
the matrix element of $Q_{12}$ remains vanishing.

As a consequence of the previous remarks,
our results for the individual contributions of the operators $Q_{1-10}$
to $\varepsilon'/\varepsilon$ may differ slightly from those
reported in ref.~\cite{Monaco}.
We have however checked that, in the ten-operator case, we reproduce
 their NLO results exactly.

Tables 7, 8 and 9 show the contributions to
$\varepsilon'/\varepsilon$ of each operator, for different choices of
$\Lambda_{QCD}^{(4)}$ and $m_t$, in the HV scheme.
The first ten contributions are also partially grouped in a ``positive''
gluonic component versus a ``negative'' electroweak component, which
shows
the ``superweak'' behavior of $\varepsilon'/\varepsilon$ within
the standard model as the top mass increases.
The total effect in the extended operator basis is given
in the last row.

We find these tables a useful way of displaying the currently available
theoretical information on $\varepsilon'/\varepsilon$. In particular
we observe that for $m_t\gtap 170$ GeV the size of $\varepsilon'/\varepsilon$
becomes comparable to the $Q_{11}$ contribution alone, signalling
the relevance
of NLO contributions to the hadronic matrix elements.

\vspace{2cm}
\bigskip
{\sc Acknowledgements}
\bigskip

We thank J. Bijnens, M. Jamin,
 A. Manohar, G. Martinelli, S. Narison, S. Peris,
A. Pich and L. Silvestrini for discussions.

M.F. thanks the ITP at Santa Barbara; his work was partially supported by NSF
Grant No. PHY89-04035. M.F. and J.O.E. thank SISSA (Trieste) for the
hospitality as this work was in progress.

 \appendix
\section{Feynman rules}

In this appendix we collect some formulas that are useful in computing the
hadronic matrix elements.

The free propagator for the constituent quark is given by
\beq
S_0 (p) = \frac{i}{\not p - M}\, ,
\eeq
where $\not p = \gamma \cdot p$.
The same propagator in the external gluon field (fixed-point gauge)
 is~\cite{S2}:
\beq
S_1 (p) = -\frac{i\ g_s}{4}\ G_{\mu\nu}^a t^a\
\frac{R^{\mu\nu}}{(p^2 - M^2)^2} \, ,
\eeq
where
\beq
R^{\mu\nu} = \sigma^{\mu\nu}(\not p + M) +
(\not p + M)\sigma^{\mu\nu}
\eeq
and $\sigma_{\mu\nu} = (i/2) [\gamma_\mu ,\,
\gamma_\nu]$.
In order to compute the gluon condensate,
the quark propagator in two external gluon fields is needed:
\beq
S_2(p) =  -\frac{i\ g_s^2}{4}\ G_{\alpha\beta}^a t^a\ G_{\mu\nu}^b t^b\
\frac{(\not p +M)(f^{\alpha\beta\mu\nu} + f^{\alpha\mu\beta\nu} +
f^{\alpha\mu\nu\beta})(\not p +M)}{(p^2 - M^2)^5} \, ,
\eeq
where
\beq
f_{\alpha\beta\mu\nu} = \gamma_\alpha (\not p +M)\gamma_\beta (\not p +M)
\gamma_\mu (\not p +M)\gamma_\nu (\not p +M)\ .
\eeq

Other useful formulas are:
\beq
\Tr g_s^2 t^a t^b G_{\mu\nu}^a G_{\alpha\beta}^b =
\frac{\pi^2}{6} \vev{\frac{\alpha_s}{\pi} GG} \left(
\delta_{\mu\alpha}\delta_{\nu\beta} - \delta_{\mu\beta}\delta_{\nu\alpha}
\right) \, ,
\eeq
and
\beq
\sigma^{\mu\nu} \sigma_{\mu\nu} = 12\ \mbox{\bf I}\, ;
\qquad \sigma^{\mu\nu}
{\gamma}_\rho \sigma_{\mu\nu} = 0 \, .
\eeq

The relevant meson--quark interactions are derived from
the lagrangian in \eq{4}, which  we write here as
\beq
{\cal L}_{\chi \mbox{\scriptsize QM}} = - M \bar{q} q + 2 i\,
\frac{M}{f} \bar{q} \gamma_5 \, \Pi \, q
+ 2\, \frac{M}{f^2} \bar{q}\, \Pi^2 q + O (1/f^3 ) \, ,
\eeq
where
\beq
\Pi = \frac{1}{2} \sum_a \lambda^a \pi^a = \frac{1}{\sqrt{2}}
\left[ \begin{array}{ccc} \tilde{\pi}^0 & \pi^+ & K^+ \\
                          \pi^- & -\bar{\pi}^0 & K^0 \\
                           K^- & \bar{K}^0 & \tilde{\pi}^8 \end{array}
\right]  \, ,
\eeq
and
\beq
\tilde{\pi}^0 = \frac{1}{\sqrt{2}} \pi^0 + \frac{1}{\sqrt{6}} \eta_8\, , \qquad
\bar{\pi}^0 = \frac{1}{\sqrt{2}} \pi^0 -
\frac{1}{\sqrt{6}} \eta_8\, , \qquad \tilde{\pi}^8 = -
\frac{2}{\sqrt{6}} \eta_8 \, ,
\eeq
which yields
\beq
\bar{q} \gamma_5 \, \Pi \, q = \frac{1}{\sqrt{2}} \left( \bar{u} \gamma_5 u
\, \tilde{\pi}^0 - \bar{d} \gamma_5 d \, \bar{\pi}^0 + \bar{d} \gamma_5 s \,
K^0
+ \bar{u} \gamma_5 d \, \pi^+ + \cdots \: \right) \, .
\eeq

The relevant Feynman rules are therefore:
\bea
K^0\   \bar d\gamma_5 s -\mbox{coupling:}
&\qquad\qquad & - \frac{M\sqrt{2}}{f} \gamma_5 \\
K^+\   \bar u\gamma_5 s -\mbox{coupling:}
&\qquad\qquad & - \frac{M\sqrt{2}}{f} \gamma_5 \\
\pi^0\ \bar d\gamma_5 d -\mbox{coupling:}
&\qquad\qquad & + \frac{M}{f} \gamma_5 \\
\pi^0\ \bar u\gamma_5 u -\mbox{coupling:}
&\qquad\qquad & - \frac{M}{f} \gamma_5 \\
\pi^+\ \bar u\gamma_5 d -\mbox{coupling:}
&\qquad\qquad & - \frac{M\sqrt{2}}{f} \gamma_5 \\
K^0\ \pi^0\ \bar d s     -\mbox{coupling:}
&\qquad\qquad & - i\frac{M}{f^2\sqrt{2}} \\
K^0\ \pi^+\ \bar u s     -\mbox{coupling:}
&\qquad\qquad & + i\frac{M}{f^2} \\
K^+\ \pi^0\ \bar u s     -\mbox{coupling:}
&\qquad\qquad & + i\frac{M}{f^2\sqrt{2}} \\
K^+\ \pi^-\ \bar d s     -\mbox{coupling:}
&\qquad\qquad & + i\frac{M}{f^2} \, .
\eea
All meson fields are entering the vertex.
The same rules hold for the conjugate couplings.

%--------------------- TABLES --------------------------------------
\clearpage
\begin{table}
\begin{center}
\begin{tabular}{|c||c|c|c|c|c|c|}
%\hline
%\multicolumn{6}{|c|}{$\vev{\overline{q}q}_G$}\\
\hline
 $M$ (MeV)    &  120  & 140  &  160  &  180  &  200   &  220   \\
\hline
&   &   &  &  &   &  \\
$\vev{\overline{q}q}_G^{1/3}$ (MeV)
           & $-315$ & $-299$ & $-286$ & $-275$ & $-265$ & $-257$    \\
&   &   &  &  &   & \\
\hline
\hline
 $M$ (MeV)    &  240  & 260  &  280  &  300  &  320   &  340   \\
\hline
&   &   &  &  &   &  \\
$\vev{\overline{q}q}_G^{1/3}$ (MeV)
           & $-250$ & $-243$ & $-237$ & $-232$ & $-227$ & $-222$    \\
&   &   &  &  &   & \\
\hline
\end{tabular}
\end{center}
\caption{Values of $\vev{\overline{q}q}_G =
 - \frac{1}{12M} \, \langle \frac{\alpha_s}{\pi}G G \rangle$
as a function of $M$.
The central value
$\langle \frac{\alpha_s}{\pi}G G \rangle = (460\ \mbox{MeV})^4$ has been used.
The scaling of the given entries for other values of the gluon condensate
is straightforward.}
\end{table}
\begin{table}
\begin{small}
\begin{center}
\begin{tabular}{|c||c|c||c|c|}
\hline
\multicolumn{5}{|c|}{$L_5\ \times 10^{3}$ }\\
%\hline
\multicolumn{5}{|c|}{
 ($\mu=1.4$ GeV,\ \ $\vev{\overline{q}q}^{1/3} = - 273$ MeV) }\\
\hline
$\Lambda_\chi$ (GeV)& \multicolumn{2}{c||}{0.83}
               & \multicolumn{2}{c|}{1.0} \\
\hline
$M$ (MeV) & NDR & HV & NDR & HV \\
\hline
120 &3.4 &3.7 &3.6 &3.7  \\
140 &2.8 &3.1 &2.9 &3.2  \\
160 &2.3 &2.6 &2.5 &2.7  \\
180 &1.9 &2.2 &2.1 &2.4  \\
200 &1.5 &1.9 &1.8 &2.1  \\
220 &1.2 &1.6 &1.5 &1.8  \\
240 &1.0 &1.4 &1.3 &1.6  \\
260 &0.7 &1.2 &1.1 &1.4  \\
280 &0.5 &1.0 &0.9 &1.2  \\
300 &0.3 &0.8 &0.7 &1.1  \\
320 &0.2 &0.7 &0.6 &0.9  \\
340 &0.0 &0.5 &0.4 &0.8  \\
\hline
\end{tabular}
\hspace{.5cm}
\begin{tabular}{|c||c|c||c|c|}
\hline
\multicolumn{5}{|c|}{$L_5\ \times 10^{3}$ }\\
%\hline
\multicolumn{5}{|c|}{
 ($\mu=1.0$ GeV,\ \ $\vev{\overline{q}q}^{1/3} = - 259$ MeV) }\\
\hline
$\Lambda_\chi$ (GeV)& \multicolumn{2}{c||}{0.83}
               & \multicolumn{2}{c|}{1.0} \\
\hline
$M$ (MeV) & NDR & HV & NDR & HV \\
\hline
120 &4.0 &4.3 &4.2 &4.4  \\
140 &3.2 &3.6 &3.5 &3.7  \\
160 &2.7 &3.0 &2.9 &3.2  \\
180 &2.2 &2.6 &2.5 &2.7  \\
200 &1.8 &2.2 &2.1 &2.4  \\
220 &1.4 &1.9 &1.8 &2.1  \\
240 &1.1 &1.6 &1.5 &1.8  \\
260 &0.9 &1.4 &1.3 &1.6  \\
280 &0.6 &1.1 &1.0 &1.4  \\
300 &0.4 &0.9 &0.8 &1.2  \\
320 &0.2 &0.7 &0.7 &1.0  \\
340 &0.0 &0.5 &0.5 &0.9  \\
\hline
\end{tabular}
\end{center}
\end{small}
\caption{Table of $L_5(\mu)$
obtained in the $\chi$QM for different values of $M$. The results
are shown for $\Lambda_{QCD}^{(4)} = 300$ MeV. }
\end{table}
\begin{table}
\begin{center}
\begin{tabular}{|c||r|r|r|r|r|r|}
\hline
 $M$ (MeV)    &  120  & 140  &  160  &  180  &  200   &  220   \\
\hline
\hline
&   &   &  &  &   &  \\
$\displaystyle{\left|
\frac{\vev{Q_{11}}^{NLO}}{\vev{Q_{6}}^{LO}_{NDR}}\right| }$
& 1\ \% & $1\ \%$  & $ 1\ \%$ & $1\ \%$ & $2\ \%$ & $2\ \%$    \\
& &  &  &  & & \\
\hline
&  &   & &  &  & \\
$\displaystyle{\left|
\frac{\vev{Q_{11}}^{NLO}}{\vev{Q_{6}}^{LO}_{HV}}\right|}$
& 1\ \% & $1\ \% $ & $ 1\ \%$ & $1\ \%$ & $2\ \%$ & $2\ \%$   \\
&  &   &  &  &   &  \\
\hline
\hline
&  &  &  &  &  & \\
$\displaystyle{\left|
\frac{\vev{Q_{6}}^{NLO}_{NDR}}{\vev{Q_{6}}^{LO}_{NDR}}\right|}$
& 14\ \%  & $15\ \% $  & $17\ \%$ & $20\ \%$ & $22\ \% $ & $25\ \%$   \\
&  &  &  &  &  &  \\
\hline
&   &   &  &  &   &  \\
$\displaystyle{\left|
\frac{\vev{Q_6}^{NLO}_{HV}}{\vev{Q_{6}}^{LO}_{HV}}\right| }$
& 14\ \% & $16\ \%$  & $18\ \%$ & $20\ \%$ & $22\ \%$ & $25\ \%$    \\
& &  &  &  & & \\
\hline
\end{tabular}
\end{center}
\caption{Typical weights of NLO contributions to the
$K^0\to\pi^+\pi^-$ amplitude
in the $\chi$QM. The results are shown for $\mu= 1$ GeV,
$\Lambda^{(4)}_{QCD} = 300$ MeV, $\Lambda_\chi = 1.0$ GeV and
$m_t = 170$ GeV, and include the NLO Wilson coefficients (``LO'' for
$Q_{11}$).}
\end{table}
\clearpage
\begin{table}
\begin{small}
\begin{center}
\begin{tabular}{|c||c|c|r||c|c|r||c|c|r|}
\hline
\multicolumn{10}{|c|}{``$\varepsilon'/\varepsilon$ '' $\times 10^{4} \qquad$
 ($\mu=1.4$ GeV, $\Lambda_\chi = 0.83$ GeV)}\\
\hline
$\Lambda_{QCD}^{(4)}$ & \multicolumn{3}{c||}{ 200 MeV}
& \multicolumn{3}{c||}{300 MeV}
&\multicolumn{3}{c|}{400 MeV} \\
\hline
\hline
 & NDR & HV & $\Delta$ & NDR & HV & $\Delta$ & NDR & HV & $\Delta$ \\
 \hline
 $1/N_c$    & 8.7 & 7.3 & $- 19 \ \%$ & 11 & 9.0 & $- 22 \ \%$ & 14 & 11 &
 $- 26 \ \% $ \\
\hline
\hline
$M$(MeV) & NDR & HV & $\Delta$ & NDR & HV & $\Delta$ & NDR & HV & $\Delta$ \\
\hline
120 & 21 & 21 & $- 1 \ \%$ & 27 &  26   &  $- 6\ \%$ & 34  & 32 &
$- 6 \ \% $ \\
140 & 15  & 16  &  $ 4 \ \%$ & 19 & 20 & $ 2 \ \%$ & 24  & 24 &
$- 0 \ \% $\\
160 & 10  & 12 & 14 \% & 13  & 15 & $ 12 \ \% $ & 16  & 18 & $ 9 \ \% $ \\
180 & 6.5 & 9.2 & 30 \% & 8.1  & 11 & 28 \% & 10 & 14 & $ 26\ \%$ \\
200 & 3.2 & 7.0 & 54 \%  & 4.0 & 8.6 & 54 \%  & 5.0 & 10 & 52\ \% \\
220 & 0.4 & 5.4 & 92 \% & 0.4 & 6.5 & 94 \% & 0.5 & 7.9 & 94 \% \\
\hline
\end{tabular}
\vspace{1cm}\\
\begin{tabular}{|c||c|c|r||c|c|r||c|c|r|}
\hline
\multicolumn{10}{|c|}{``$\varepsilon'/\varepsilon$ ''$ \times 10^{4}\qquad$
 ($\mu=1.0$ GeV, $\Lambda_\chi = 0.83$ GeV)}\\
\hline
$\Lambda_{QCD}^{(4)}$ & \multicolumn{3}{c||}{ 200 MeV}
& \multicolumn{3}{c||}{300 MeV}
&\multicolumn{3}{c|}{400 MeV} \\
\hline
\hline
 & NDR & HV & $\Delta$ & NDR & HV & $\Delta$ & NDR & HV & $\Delta$ \\
 \hline
 $1/N_c$    & 8.0 & 6.5 & $- 24 \ \%$ & 10 & 7.6 & $- 33 \ \%$ & 13 & 8.7 &
 $- 46 \ \% $ \\
\hline
\hline
$M$(MeV) & NDR & HV & $\Delta$ & NDR & HV & $\Delta$ & NDR & HV & $\Delta$ \\
\hline
120 & 23 & 22 & $- 4 \ \%$ & 31 &  28   & $- 10 \ \%$ & 43  & 37 &
$- 18 \ \%$ \\
140 & 17  & 17  &  1 \% & 23 & 22 & $- 5 \ \%$ & 32   & 28  &
$- 14 \ \% $ \\
160 & 12  & 13 & 11 \% & 16 & 17 & 4 \%  & 23  & 22 & $- 6\ \%$  \\
180 & 7.5 & 10 & 25 \% & 10  & 13 & 17 \% & 16 & 17 & 7 \% \\
200 & 4.1 & 7.8 & 47 \%  & 6.3 & 10 & 38 \%  & 11 & 14 & 26 \% \\
220 & 1.1 & 6.1 & 81 \% & 2.4 & 8.1 & 70 \% & 5.1 & 11 & 54 \% \\
\hline
\end{tabular}
\end{center}
\end{small}
\caption{Toy model for the
$\gamma_5$-scheme dependence of $\varepsilon '/\varepsilon$.
The results are shown for $m_t = 170$ GeV,
$\Lambda_\chi=0.83$ GeV, and $\delta_{KM}$ in the first quadrant.
The two tables refer to the choice $\mu= m_c = 1.4$ GeV and 1.0 GeV
respectively for the renormalization scale.
The $\chi$QM results are compared
with the $1/N_c$ predictions evaluated according to ref.~[2].
The parameter $\Delta$ is defined as the (HV $-$ NDR)/HV combination
of the entries. }
\end{table}
\clearpage
\begin{table}
\begin{small}
\begin{center}
\begin{tabular}{|c||c|c|r||c|c|r||c|c|r|}
\hline
\multicolumn{10}{|c|}{``$\varepsilon'/\varepsilon$ ''$ \times 10^{4} \qquad$
 ($\mu=1.4$ GeV, $\Lambda_\chi = 1.0$ GeV)}\\
\hline
$\Lambda_{QCD}^{(4)}$ & \multicolumn{3}{c||}{ 200 MeV}
& \multicolumn{3}{c||}{300 MeV}
&\multicolumn{3}{c|}{400 MeV} \\
\hline
\hline
 & NDR & HV & $\Delta$ & NDR & HV & $\Delta$ & NDR & HV & $\Delta$ \\
 \hline
 $1/N_c$    & 4.4 & 3.3 & $- 33 \ \%$ & 5.5 & 4.0 & $- 37 \ \%$ & 7.0 & 4.9 &
 $- 43 \ \% $ \\
\hline
\hline
$M$(MeV) & NDR & HV & $\Delta$ & NDR & HV & $\Delta$ & NDR & HV & $\Delta$ \\
\hline
120 & 23 & 21 & $- 6 \ \%$ & 29  &  26   & $- 8 \ \%$ & 36  & 32 &
$- 11 \ \%$ \\
140 & 17  & 16  & $- 3 \ \%$ & 21 & 20 & $- 5 \ \%$ & 27 & 25 &
$- 8 \ \%$ \\
160 & 12  & 13 & $ 2 \ \%$ & 15 & 15 & $- 0 \ \%$  & 19 & 19 &
$- 3 \ \% $ \\
180 & 8.7 & 9.7 & 11 \% & 11  & 12 & $ 9 \ \%$ & 14 & 14 & $ 5 \ \%$ \\
200 & 5.6 & 7.5 & 24 \%  & 7.1 & 9.1 & 23 \%  & 8.9 & 11 & $20\ \%$ \\
220 & 3.1 & 5.7 & 46 \% & 3.8 & 7.0 & 46 \% & 4.8 & 8.4 & 43 \% \\
\hline
\end{tabular}
\vspace{1cm}\\
\begin{tabular}{|c||c|c|r||c|c|r||c|c|r|}
\hline
\multicolumn{10}{|c|}{``$\varepsilon'/\varepsilon$ ''$ \times 10^{4}\qquad $
 ($\mu=1.0$ GeV, $\Lambda_\chi = 1.0$ GeV)}\\
\hline
$\Lambda_{QCD}^{(4)}$ & \multicolumn{3}{c||}{ 200 MeV}
& \multicolumn{3}{c||}{300 MeV}
&\multicolumn{3}{c|}{400 MeV} \\
\hline
\hline
 & NDR & HV & $\Delta$ & NDR & HV & $\Delta$ & NDR & HV & $\Delta$ \\
 \hline
 $1/N_c$    & 4.0 & 2.7 & $- 45 \ \%$ & 5.0 & 3.1 & $- 60 \ \%$ & 6.4 & 3.5 &
$- 84 \ \%$ \\
\hline
\hline
$M$(MeV) & NDR & HV & $\Delta$ & NDR & HV & $\Delta$ & NDR & HV & $\Delta$ \\
\hline
120 & 25 & 23 & $- 9 \ \%$ & 33 &  29   & $- 15 \ \%$ & 46  & 38 &
$- 23 \ \%$ \\
140 & 18  & 17 & $- 6 \ \%$ & 25 & 22 & $- 12 \ \%$ & 35   & 29  &
$- 21 \ \%$ \\
160 & 14  & 13 & $- 1 \ \%$ & 19  & 17 & $- 8 \ \%$  & 27  & 23 &
$- 17 \ \%$  \\
180 & 9.8 & 10 & 7 \% & 14  & 14 & $-1 \%$ & 20 & 18 & $- 11 \ \%$ \\
200 & 6.6 & 8.2 & 19 \%  & 9.7 & 11 & 11 \%  & 15 & 15 & $- 1\ \%$  \\
220 & 3.9 & 6.5 & 39 \% & 6.1 & 8.6 & 29 \% & 10 & 12 & 15 \% \\
\hline
\end{tabular}
\end{center}
\end{small}
\caption{Same as Table 2, with $\Lambda_\chi=1$ GeV.}
\end{table}
\clearpage
\begin{table}
%\begin{footnotesize}
\begin{center}
\begin{tabular}{|c|r r||r r||r r|}
\hline
$\Lambda_{QCD}^{(4)}$ & \multicolumn{2}{c||}{ 200 MeV }
                      & \multicolumn{2}{c||}{ 300 MeV }
                      & \multicolumn{2}{c| }{ 400 MeV } \\
%\hline
%$m_t$&\multicolumn{2}{c|}{$130$ GeV}&\multicolumn{2}{c||}{$170$ GeV}
%     &\multicolumn{2}{c|}{$130$ GeV}&\multicolumn{2}{c|}{$170$ GeV}  \\
\hline \hline
$z_1$&$(0.031)$&$-0.477$&$(0.033)$&$-0.606$&$(0.035)$&$-0.774$ \\
\hline
$z_2$&$(0.988)$&$1.256$&$(0.988)$&$1.346$&$(0.987)$&$1.472$ \\
\hline
$z_3$&$  $&$0.005$&$  $&$0.008$&$  $&$0.015$ \\
\hline
$z_4$&$  $&$-0.011$&$  $&$-0.018$&$  $&$-0.030$ \\
\hline
$z_5$&$  $&$0.003$&$  $&$0.004$&$  $&$0.006$ \\
\hline
$z_6$&$  $&$-0.010$&$  $&$-0.016$&$  $&$-0.026$ \\
\hline
$z_7/\alpha$&$  $&$-0.005$&$  $&$-0.003$&$  $&$-0.002$ \\
\hline
$z_8/\alpha$&$  $&$0.007$&$  $&$0.011$&$  $&$0.018$ \\
\hline
$z_9/\alpha$&$  $&$-0.001$&$  $&$0.004$&$  $&$0.010$ \\
\hline
$z_{10}/\alpha$&$  $&$-0.007$&$  $&$-0.011$&$  $&$-0.017$ \\
\hline
$z_{11}$&$  $&$-0.035$&$  $&$-0.044$&$  $&$-0.056$ \\
\hline
$z_{12}$&$  $&$0.342$&$  $&$0.486$&$  $&$0.690$ \\
\hline
\hline
$y_3$&$(0.001)$&$0.026$&$(0.001)$&$0.034$&$(0.000)$&$0.044$ \\
\hline
$y_4$&$(0.001)$&$-0.046$&$(0.001)$&$-0.056$&$(0.001)$&$-0.066$ \\
\hline
$y_5$&$(0.000)$&$0.013$&$(0.000)$&$0.015$&$(0.000)$&$0.018$ \\
\hline
$y_6$&$(0.001)$&$-0.066$&$(0.001)$&$-0.088$&$(0.001)$&$-0.120$ \\
\hline
$y_7/\alpha$&$(0.151)$&$-0.031$&$(0.151)$&$-0.029$&$(0.151)$&$-0.027$ \\
\hline
$y_8/\alpha$&$(0.000)$&$0.126$&$(0.000)$&$0.172$&$(0.000)$&$0.240$ \\
\hline
$y_9/\alpha$&$(-1.094)$&$-1.541$&$(-1.094)$&$-1.632$&$(-1.094)$&$-1.759$ \\
\hline
$y_{10}/\alpha$&$(0.000)$&$0.560$&$(0.000)$&$0.703$&$(0.000)$&$0.888$ \\
\hline
$y_{11}$&$(-0.193)$&$-0.343$&$(-0.193)$&$-0.371$&$(-0.193)$&$-0.414$ \\
\hline
$y_{12}$&$(1.158)$&$2.144$&$(1.158)$&$2.296$&$(1.158)$&$2.482$ \\
\hline
\end{tabular}
\end{center}
%\end{footnotesize}
\caption{NLO Wilson coefficients at $\mu=1$ GeV in the HV
scheme for $m_t=170$~GeV ($\alpha=1/128$).
The corresponding values at $\mu=m_W$ are given in parenthesis.
In addition, at $\mu=m_c$ we have $z_{3-12}(m_c)=0$. The coefficients
$y_{11}$ and $y_{12}$ are given at the ``LO'' (QCD two-loop mixing)
and their values are $\gamma_5$-scheme independent.}
\end{table}
\clearpage
\begin{table}
\begin{footnotesize}
\begin{center}
\begin{tabular}{|c||r|r|r|r||r|r|r|r||r|r|r|r|}
\hline
\multicolumn{13}{|c|}{$\varepsilon'/\varepsilon\times 10^{4}$\ \ (NLO)} \\
\hline
$m_t$ &\multicolumn{4}{c||}{ $150$ GeV} & \multicolumn{4}{c||}{ $170$ GeV}
      &\multicolumn{4}{c|}{ $190$ GeV}
 \\ \hline
$\delta_{KM}$&\multicolumn{2}{c|}{I quad.}&\multicolumn{2}{c||}{II quad.}
             &\multicolumn{2}{c|}{I quad.}&\multicolumn{2}{c||}{II quad.}
             &\multicolumn{2}{c|}{I quad.}&\multicolumn{2}{c|}{II quad.}
 \\ \hline \hline
$Q_3$ &$0.3$&  &$0.2$&  &$0.3$&  &$0.2$&  &$0.3$&  &$0.2$& \\
\cline{1-2}\cline{4-4}\cline{6-6}\cline{8-8}\cline{10-10}\cline{12-12}
$Q_4$ &$-4.7$&  &$-3.3$&  &$-4.4$&  &$-3.0$&  &$-4.0$&  &$-2.7$& \\
\cline{1-2}\cline{4-4}\cline{6-6}\cline{8-8}\cline{10-10}\cline{12-12}
$Q_5$
&$-0.6$&$4.1$&$-0.4$&$2.9$&$-0.5$&$3.8$&$-0.4$&$2.6$&$-0.5$&$3.5$&$-0.3$&$2.
3$ \\
\cline{1-2}\cline{4-4}\cline{6-6}\cline{8-8}\cline{10-10}\cline{12-12}
$Q_6$ &$9.1$&  &$6.3$&  &$8.4$&  &$5.6$&  &$7.8$&  &$5.1$& \\
\hline
$Q_7$ &$0.7$&  &$0.5$&  &$0.4$&  &$0.2$&  &$0.0$&  &$0.0$& \\
\cline{1-2}\cline{4-4}\cline{6-6}\cline{8-8}\cline{10-10}\cline{12-12}
$Q_8$ &$-4.7$&  &$-3.3$&  &$-5.3$&  &$-3.6$&  &$-6.0$&  &$-3.9$& \\
\cline{1-2}\cline{4-4}\cline{6-6}\cline{8-8}\cline{10-10}\cline{12-12}
$Q_9$
&$2.9$&$-1.8$&$2.0$&$-1.2$&$3.1$&$-2.7$&$2.1$&$-1.8$&$3.2$&$-3.6$&$2.1$&$-2.
4$ \\
\cline{1-2}\cline{4-4}\cline{6-6}\cline{8-8}\cline{10-10}\cline{12-12}
$Q_{10}$ &$-0.7$&  &$-0.5$&  &$-0.8$&  &$-0.5$&  &$-0.8$&  &$-0.5$& \\
\hline
$Q_{11}$ &\multicolumn{2}{r|}{$-0.4$} & \multicolumn{2}{r||}{$-0.3$}
         &\multicolumn{2}{r|}{$-0.3$} & \multicolumn{2}{r||}{$-0.2$}
         &\multicolumn{2}{r|}{$-0.3$} & \multicolumn{2}{r|}{$-0.2$}
 \\ \hline \hline
 All    &\multicolumn{2}{r|}{$2.0$} & \multicolumn{2}{r||}{$1.4$}
         &\multicolumn{2}{r|}{$0.8$} & \multicolumn{2}{r||}{$0.5$}
         &\multicolumn{2}{r|}{$-0.4$} & \multicolumn{2}{r|}{$-0.2$}
\\ \hline
%Ref.~\cite{Monaco} & \multicolumn{2}{r||}{$$} & \multicolumn{2}{r||}{$$}
%                  & \multicolumn{2}{r||}{$$} & \multicolumn{2}{r|}{$$}
%\\ \hline
\end{tabular}\end{center}
\end{footnotesize}
\caption{ Anatomy of $\varepsilon'/\varepsilon$ for
$\Lambda_{QCD}^{(4)} = 200$ MeV ($\alpha_s(m_Z)_{\overline{MS}}=0.109$),
$B^{(1/2)}_6 =  B^{(3/2)}_8 = B^{(1/2)}_{11} =1$, in the HV scheme.
The contribution of each operator is shown at $\mu=1$ GeV,
together with partial
grouping of the gluonic and the electroweak sectors.
The contribution of $Q_{12}$ is being neglected.
The entries for $Q_3-Q_{10}$ represent the $1/N_c$ central values
(for a discussion on the error bars see  ref. [2],
whose input parameters have been thoroughly assumed).
The $Q_{11}$ amplitude is computed in the $\chi$QM, using the
``LO'' Wilson coefficient and setting
$\vev{\overline{q}q}_G$ $=$ $-(275\ \mbox{MeV})^3$
as a typical value
(see Table 1).}
\end{table}
\clearpage
\begin{table}
\begin{footnotesize}
\begin{center}
\begin{tabular}{|c||r|r|r|r||r|r|r|r||r|r|r|r|}
\hline
\multicolumn{13}{|c|}{$\varepsilon'/\varepsilon\times 10^{4}$\ \ (NLO)} \\
\hline
$m_t$ &\multicolumn{4}{c||}{ $150$ GeV} & \multicolumn{4}{c||}{ $170$ GeV}
      &\multicolumn{4}{c|}{ $190$ GeV}
 \\ \hline
$\delta_{KM}$&\multicolumn{2}{c|}{I quad.}&\multicolumn{2}{c||}{II quad.}
             &\multicolumn{2}{c|}{I quad.}&\multicolumn{2}{c||}{II quad.}
             &\multicolumn{2}{c|}{I quad.}&\multicolumn{2}{c|}{II quad.}
 \\ \hline \hline
$Q_3$ &$0.5$&  &$0.3$&  &$0.5$&  &$0.3$&  &$0.4$&  &$0.3$& \\
\cline{1-2}\cline{4-4}\cline{6-6}\cline{8-8}\cline{10-10}\cline{12-12}
$Q_4$ &$-5.0$&  &$-3.4$&  &$-4.6$&  &$-3.1$&  &$-4.2$&  &$-2.8$& \\
\cline{1-2}\cline{4-4}\cline{6-6}\cline{8-8}\cline{10-10}\cline{12-12}
$Q_5$
&$-0.6$&$6.2$&$-0.5$&$4.3$&$-0.6$&$5.7$&$-0.4$&$3.8$&$-0.5$&$5.3$&$-0.4$&$3.
5$ \\
\cline{1-2}\cline{4-4}\cline{6-6}\cline{8-8}\cline{10-10}\cline{12-12}
$Q_6$ &$11.3$&  &$7.8$&  &$10.4$&  &$7.0$&  &$9.6$&  &$6.4$& \\
\hline
$Q_7$ &$0.6$&  &$0.4$&  &$0.3$&  &$0.2$&  &$0.0$&  &$0.0$& \\
\cline{1-2}\cline{4-4}\cline{6-6}\cline{8-8}\cline{10-10}\cline{12-12}
$Q_8$ &$-5.9$&  &$-4.1$&  &$-6.7$&  &$-4.5$&  &$-7.5$&  &$-5.0$& \\
\cline{1-2}\cline{4-4}\cline{6-6}\cline{8-8}\cline{10-10}\cline{12-12}
$Q_9$
&$3.0$&$-3.2$&$2.1$&$-2.3$&$3.1$&$-4.3$&$2.1$&$-2.9$&$3.3$&$-5.3$&$2.2$&$-3.
5$ \\
\cline{1-2}\cline{4-4}\cline{6-6}\cline{8-8}\cline{10-10}\cline{12-12}
$Q_{10}$ &$-0.9$&  &$-0.6$&  &$-1.0$&  &$-0.7$&  &$-1.0$&  &$-0.7$& \\
\hline
$Q_{11}$ &\multicolumn{2}{r|}{$-0.4$} & \multicolumn{2}{r||}{$-0.3$}
         &\multicolumn{2}{r|}{$-0.4$} & \multicolumn{2}{r||}{$-0.2$}
         &\multicolumn{2}{r|}{$-0.3$} & \multicolumn{2}{r|}{$-0.2$}
 \\ \hline \hline
 All    &\multicolumn{2}{r|}{$2.6$} & \multicolumn{2}{r||}{$1.8$}
         &\multicolumn{2}{r|}{$1.1$} & \multicolumn{2}{r||}{$0.7$}
         &\multicolumn{2}{r|}{$-0.4$} & \multicolumn{2}{r|}{$-0.2$}
\\ \hline
%Ref.~\cite{Monaco} & \multicolumn{2}{r||}{$$} & \multicolumn{2}{r||}{$$}
%                  & \multicolumn{2}{r||}{$$} & \multicolumn{2}{r|}{$$}
%\\ \hline
\end{tabular}\end{center}
\end{footnotesize}
\caption{Same as in Table 5 for $\Lambda_{QCD}^{(4)} = 300$ MeV
         ($\alpha_s(m_Z)_{\overline{MS}}=0.116$).}
\end{table}
\clearpage
\begin{table}
\begin{footnotesize}
\begin{center}
\begin{tabular}{|c||r|r|r|r||r|r|r|r||r|r|r|r|}
\hline
\multicolumn{13}{|c|}{$\varepsilon'/\varepsilon\times 10^{4}$\ \ (NLO)} \\
\hline
$m_t$ &\multicolumn{4}{c||}{ $150$ GeV} & \multicolumn{4}{c||}{ $170$ GeV}
      &\multicolumn{4}{c|}{ $190$ GeV}
 \\ \hline
$\delta_{KM}$&\multicolumn{2}{c|}{I quad.}&\multicolumn{2}{c||}{II quad.}
             &\multicolumn{2}{c|}{I quad.}&\multicolumn{2}{c||}{II quad.}
             &\multicolumn{2}{c|}{I quad.}&\multicolumn{2}{c|}{II quad.}
 \\ \hline \hline
$Q_3$ &$0.7$&  &$0.5$&  &$0.7$&  &$0.5$&  &$0.6$&  &$0.4$& \\
\cline{1-2}\cline{4-4}\cline{6-6}\cline{8-8}\cline{10-10}\cline{12-12}
$Q_4$ &$-5.2$&  &$-3.6$&  &$-4.8$&  &$-3.2$&  &$-4.4$&  &$-2.9$& \\
\cline{1-2}\cline{4-4}\cline{6-6}\cline{8-8}\cline{10-10}\cline{12-12}
$Q_5$
&$-0.7$&$8.7$&$-0.5$&$6.0$&$-0.7$&$8.0$&$-0.4$&$5.4$&$-0.6$&$7.4$&$-0.4$&$4.
9$ \\
\cline{1-2}\cline{4-4}\cline{6-6}\cline{8-8}\cline{10-10}\cline{12-12}
$Q_6$ &$13.8$&  &$9.6$&  &$12.7$&  &$8.6$&  &$11.8$&  &$7.8$& \\
\hline
$Q_7$ &$0.5$&  &$0.4$&  &$0.2$&  &$0.2$&  &$0.0$&  &$0.0$& \\
\cline{1-2}\cline{4-4}\cline{6-6}\cline{8-8}\cline{10-10}\cline{12-12}
$Q_8$ &$-7.4$&  &$-5.1$&  &$-8.4$&  &$-5.7$&  &$-9.4$&  &$-6.2$& \\
\cline{1-2}\cline{4-4}\cline{6-6}\cline{8-8}\cline{10-10}\cline{12-12}
$Q_9$
&$3.2$&$-5.0$&$2.2$&$-3.4$&$3.3$&$-6.1$&$2.3$&$-4.1$&$3.5$&$-7.3$&$2.3$&$-4.
8$ \\
\cline{1-2}\cline{4-4}\cline{6-6}\cline{8-8}\cline{10-10}\cline{12-12}
$Q_{10}$ &$-1.2$&  &$-0.9$&  &$-1.3$&  &$-0.9$&  &$-1.4$&  &$-0.9$& \\
\hline
$Q_{11}$ &\multicolumn{2}{r|}{$-0.4$} & \multicolumn{2}{r||}{$-0.3$}
         &\multicolumn{2}{r|}{$-0.4$} & \multicolumn{2}{r||}{$-0.3$}
         &\multicolumn{2}{r|}{$-0.4$} & \multicolumn{2}{r|}{$-0.3$}
 \\ \hline \hline
 All    &\multicolumn{2}{r|}{$3.3$} & \multicolumn{2}{r||}{$2.3$}
         &\multicolumn{2}{r|}{$1.5$} & \multicolumn{2}{r||}{$1.0$}
         &\multicolumn{2}{r|}{$-0.3$} & \multicolumn{2}{r|}{$-0.2$}
\\ \hline
%Ref.~\cite{Monaco} & \multicolumn{2}{r||}{$$} & \multicolumn{2}{r||}{$$}
%                  & \multicolumn{2}{r||}{$$} & \multicolumn{2}{r|}{$$}
%\\ \hline
\end{tabular}\end{center}
\end{footnotesize}
\caption{Same as in Table 5 for $\Lambda_{QCD}^{(4)} = 400$ MeV
         ($\alpha_s(m_Z)_{\overline{MS}}=0.122$).}
\end{table}
%

%----------------------- FIGURES -------------------------------
\clearpage
\begin{figure}
\epsfxsize=14cm
\centerline{\epsfbox{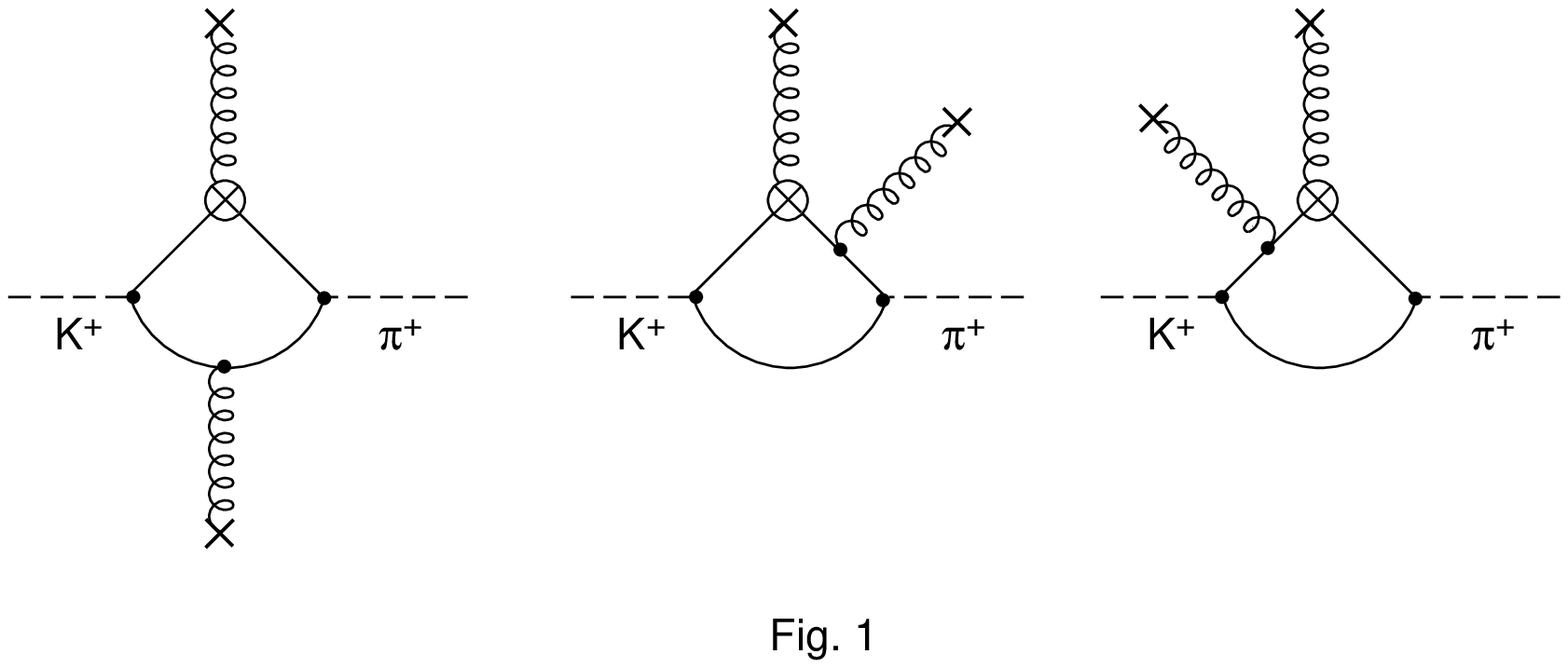}}
\caption{The relevant diagrams for the
$K^+ \rightarrow \pi^+$ transition induced by
$Q_{11}$. The circled cross indicates the operator insertion.}
\end{figure}

\begin{figure}
\epsfxsize=10cm
\centerline{\epsfbox{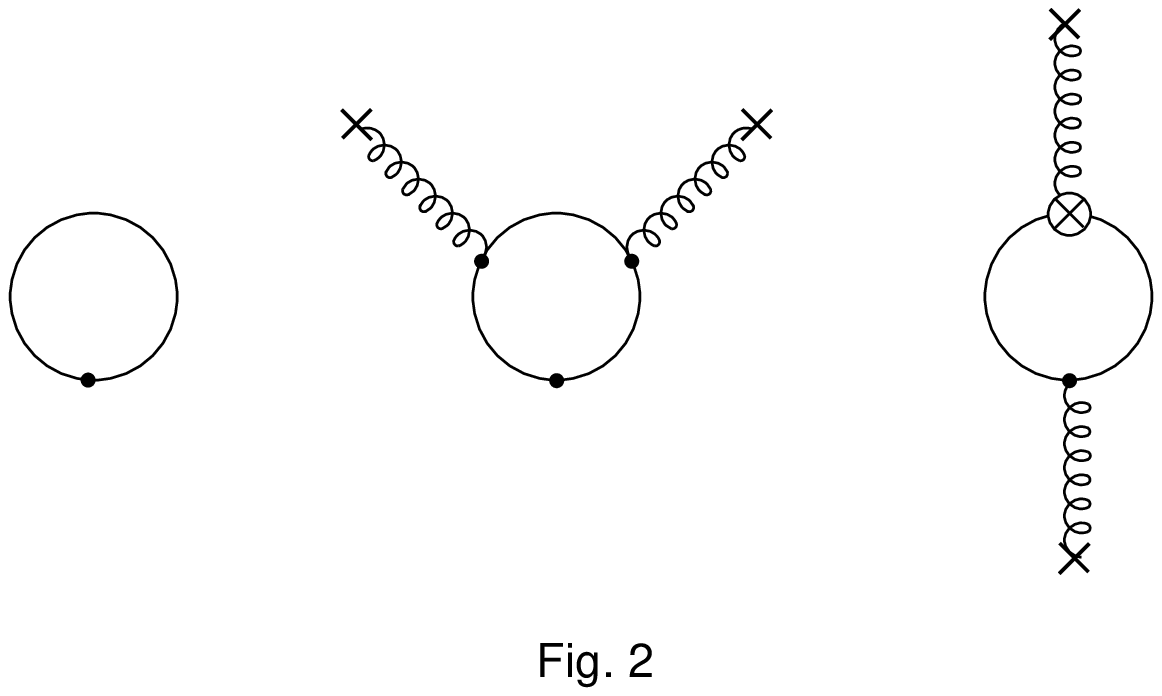}}
\caption{The diagrams for the bare quark condensate,
its  gluon-condensate correction, and the mixed condensate
$g_s\vev{\overline{q}\sigma\cdot G q}$.
The circled cross indicates the insertion of $Q_{11}$. }
\end{figure}
\clearpage
\begin{figure}
\epsfxsize=14cm
\centerline{\epsfbox{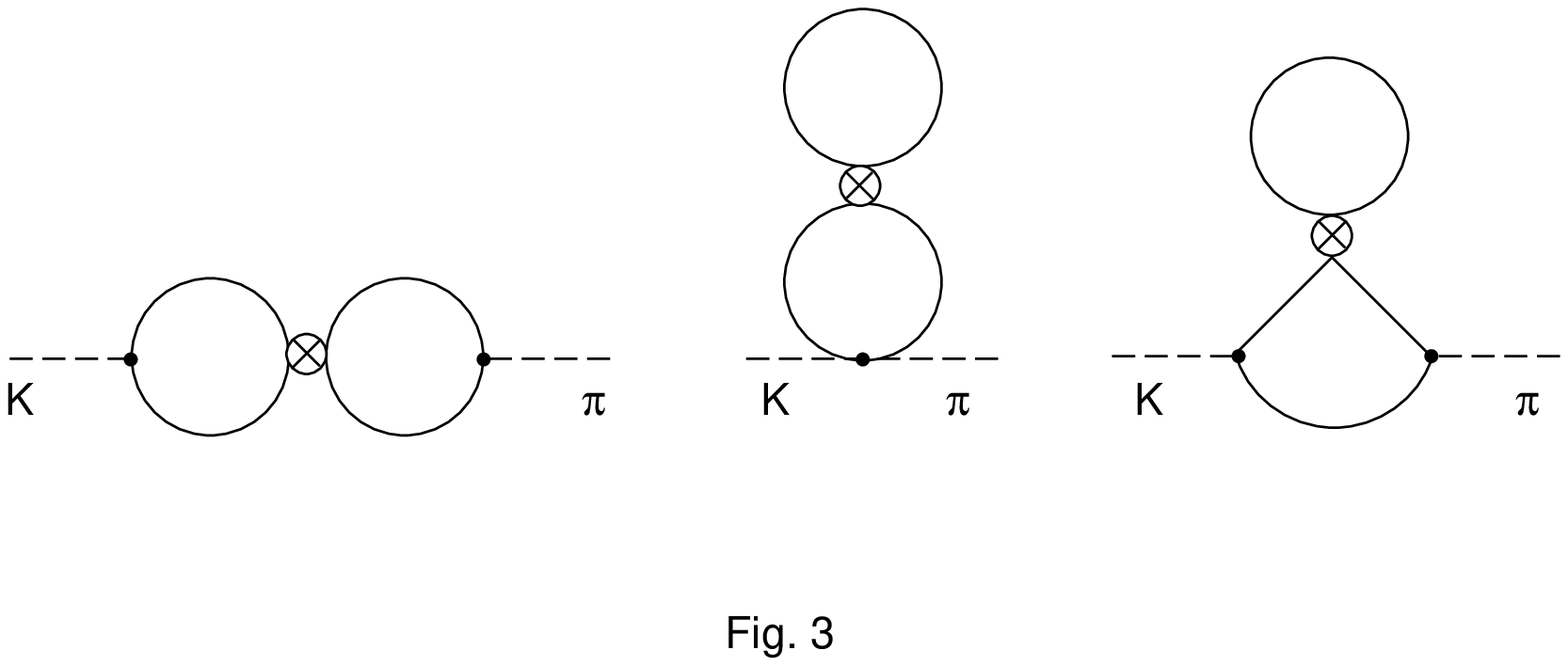}}
\caption{The relevant diagrams for the
$K \rightarrow \pi$ transition induced by
$Q_6$. The circled cross indicates the $Q_{6}$ insertion. }
\end{figure}
\begin{figure}
\epsfxsize=7.5cm
\centerline{\epsfbox{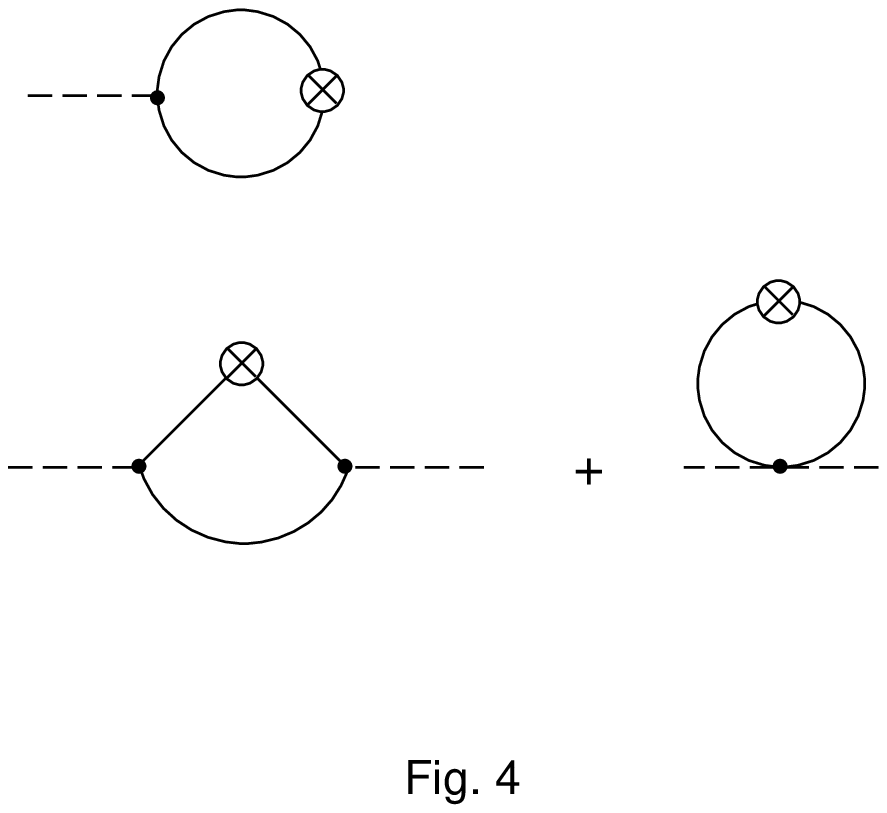}}
\caption{Building blocks for the hadronic matrix elements. The diagram
on the top corresponds to
$\vev{ 0|\,\overline{s} \gamma_5 u\, |K^+(k)}$, whereas the two diagrams
on the bottom contribute to
$\langle \pi^+(p)|\,\overline{s} d\, |K^+(k) \rangle$.
The circled cross indicates the quark current (or density) insertion. }
\end{figure}
%

%------------------------- REFERENCES ------------------------------
\clearpage
\renewcommand{\baselinestretch}{1}

\end{document}